\documentclass[aps,prb,reprint]{revtex4-2}
\usepackage{graphicx} 
\usepackage{amsmath} 
\usepackage{amssymb}
\usepackage{mathtools} 
\usepackage{gensymb} 
\usepackage{bm} 
\usepackage{natbib}
\usepackage{natmove}
\usepackage[colorlinks=true, linkcolor=blue, citecolor=blue, urlcolor=blue, linktoc=page, bookmarks=false, pdfstartview={FitH}, pdfborder={0 0 0.0 [3 3]}]{hyperref} 
\usepackage{color} 
\usepackage{dcolumn} 
\newcolumntype{.}{D{.}{.}{2.1}}
\newcolumntype{-}{D{.}{.}{4.0}}
\usepackage{makecell}
\usepackage{multirow}
\usepackage{cleveref}
\usepackage{easyReview} 
\crefname{figure}{Fig.}{Figs}
\crefname{table}{Table}{Tables}
\crefname{equation}{Eq.}{Eqs.}
\crefname{section}{Sec.}{Secs.}
\usepackage[utf8]{inputenc}
\usepackage{booktabs}
\renewcommand{\today}{\number\day \space \ifcase \month \or January\or February\or March\or April\or May\or June\or July\or August\or September\or October\or November\or December\fi \space \number\year} 
\def\m1r{\multicolumn{1}{r}}
\begin{document}
\title{Valley polarization, Rashba interaction, and weak altermagnetism in inversion-asymmetric MnPS$_\text{3}|$WS$_\text{2}$ van der Waals heterostructures}
\author{Purba \surname{Dutta}}
\email[Electronic address: ]{Purba19@iiserb.ac.in}
\author{Soumajyoti \surname{Bid}}
\author{Nirmal \surname{Ganguli}}
\email[Electronic address: ]{NGanguli@iiserb.ac.in}
\affiliation{Department of Physics, \href{https://ror.org/02rb21j89}{Indian Institute of Science Education and Research Bhopal}, Bhauri, Bhopal 462066, India}
\date{\today}
\begin{abstract}
The deliberate breaking of inversion ($\mathcal{P}$) symmetry in antiferromagnets has recently emerged as an effective means to induce various features, such as the emergence of Berry curvature, spin-valley locking, magnetoelectric coupling, and the transition from conventional antiferromagnetism to altermagnetism. Conversely, in non-magnetic systems, inversion symmetry breaking in the presence of strong spin-orbit interaction (SOI) gives rise to momentum-dependent spin splitting via the Rashba effect, enabling tunable spin polarization through external electric fields. Motivated by recent advances in two-dimensional materials, we perform first-principles calculations based on density functional theory to investigate the van der Waals (vdW) heterostructure formed by a $\mathcal{P}$-symmetric MnPS$_3$ monolayer and a WS$_2$ monolayer. We demonstrate that the interface hosts a rich interplay of emergent phenomena, including an altermagnetic phase, Rashba spin splitting, spin-valley locking, and valley polarization. Our results demonstrate that the heterostructure exhibits semiconducting behavior with a direct band gap of approximately 1.65~eV and a type-I band alignment. Remarkably, the electronic structure and band alignment can be effectively tuned between type-I and type-II regimes via an external electric field and in-plane biaxial strain. Furthermore, field-induced modulation enables strong control over the altermagnetic phase and the valley splitting. These findings establish the proposed vdW heterostructure as a highly tunable platform with significant potential for spintronic and valleytronic applications.
\end{abstract}
\maketitle
\section{\label{sec:intro}Introduction}
The emergence of two-dimensional (2D) vdW materials led to a paradigm shift in materials research by offering unprecedented ease of heterostructuring \cite{GeimN13}, besides hosting a number of highly sought-after physical properties, including high-mobility direct band-gap semiconducting nature, magnetism, spin-orbit interaction, and valley polarization \cite{WangNN12, MakPRL10, GanguliPRBR19, Ganguli2PRB19, XiaoPRL12, CaoNC12}. Owing to a weak van der Waals interaction between the layers, the heterostructures also offer the possibility of twisting the layers to a desired angle \cite{ZhongSA17, SierraNN21}. Realizing antiferromagnetism in 2D vdW materials and heterostructures presents an opportunity for realizing antiferromagnetic spintronics \cite{JungwirthNN16, BaltzRMP18, ZeleznyNP18}. Devices based on antiferromagnetic spintronics offer higher packing density due to the lack of stray fields, can operate at a THz frequency, and a Moir\'e supercell-induced flat bands of twisted bilayers \cite{LeiSA20}. Valley polarization, in which the valleys at the inequivalent high-symmetry points $K$ and $\bar{K}$ have different energies, has been observed as another key feature in 2D materials and heterostructures, leading to the proposal of valleytronic technologies \cite{SchaibleyNRM16}. Spin-valley locking may be realized in 2D materials, leading to the spin state of an electron directly tied to the valley it occupies, by combining a transition metal dichalcogenide (TMDC) with a magnetic material in a vdW heterostructure, resulting in tuning-knob control over both the spin and valley pseudospin of the electrons \cite{ZhongSA17}.

SOI in inversion-asymmetric systems often leads to a spin-momentum locking effect, known as the Rashba effect, resulting in a momentum-dependent spin splitting and a spin texture in the reciprocal space \cite{RashbaSPSS60, GanguliPRB25}. At the interfaces of certain vdW heterostructures, for example, a WS$_2$-graphene interface, the induced Rashba SOC enables the Rashba-Edelstein effect \cite{GhiasiNL19}. This phenomenon enables a standard, unpolarized electrical current to be directly converted into spin accumulation and spin current at room temperature, thereby bypassing the need to inject spins via traditional ferromagnetic electrodes. Further, the proximity effect may result in other interesting phenomena in such heterostructures, including magnetism, various types of spin-orbit interactions, and topologically nontrivial features \cite{IslandN19, TongPRAp19, QiaoPRL14}. 

The vdW layered materials have emerged as a well-suited platform for investigating two-dimensional altermagnetism \cite{HayamiPRB18, HayamiPRB20, RoojAPR23, RoojPRB25, WrzosJPCC26}. Weak interlayer vdW interactions provide a natural approach to constructing symmetry-breaking mechanisms that enable tuning of magnetic and electronic characteristics to a desired extent. Among these, the family of transition metal phosphorus trichalcogenides (TMPTs), having the chemical formula MPX$_3$ (M = Mn, Fe, Co, Ni; X = S, Se, Te), have attracted a substantial amount of attention. These materials have layered crystal structures, a strong in-plane magnetic order, and highly tunable interlayer coupling, which make them appropriate for the realization of altermagnetic phenomena through their combination of these characteristics. In addition to their magnetic richness, 2D transition-metal dichalcogenides are attractive for their valley degrees of freedom. These valley degrees of freedom couple with spin, orbital, and layer pseudospins, which provide novel opportunities in the fields of valleytronics, low-dimensional spintronics, and magneto-optoelectronic applications \cite{LiPNAS13}. The intersection of altermagnetism, symmetry engineering, and valley physics in vdW heterostructures represents a promising avenue for quantum materials research \cite{YanAPL24}. This evolving interaction provides a flexible framework for controlling spin, charge, and valley degrees of freedom, thereby enhancing our theoretical understanding of correlated quantum phenomena and paving the way for innovative information technology \cite{BurchN18}.

Monolayer WS$_2$, a prominent member of the TMDC family, has attracted significant interest due to its remarkably high carrier mobility and substantial SOI. The characteristics of these materials change substantially as they transition from bulk to monolayer, and the layering of several two-dimensional materials offers a potent and adaptable means of generating novel capabilities. Among vdW materials, these offer a pristine and manageable foundation for customizing interfacial interactions without requiring chemical or surface alterations. Specifically, vdW heterostructures that combine transition metal dichalcogenides with magnetic semiconductors like MPX$_3$ compounds have, in recent times, appeared as promising candidates for the manifestation of novel spin and valley phenomena \cite{XiaoPRL12}, influenced by stacking-induced asymmetry. A recent report on MnPS$_3|$TMDC suggests the presence of non-relativistic splitting and valley-dependent features \cite{WrzosJPCC26}.

In the present work, we design and investigate a layered 2D vertical heterojunction composed of monolayers of MnPS$_3$ and WS$_2$ using first-principles calculations. The MnPS$_3|$WS$_2$ heterostructure is predicted to exhibit unique electronic and magnetic characteristics arising from interfacial asymmetry, rendering it an intriguing candidate for spintronic applications. Monolayer WS$_2$ in the 2H phase has already been successfully synthesized experimentally \cite{GaoNC15}. The MnPS$_3$ monolayer consists of Mn$^{2+}$ ions forming a honeycomb lattice, each octahedrally coordinated by sulfur atoms belonging to adjacent (P$_2$S$_6$) ligands and orders antiferromagnetically below its N\'eel temperature $T_N = 78 \text{ K}$ \cite{WildesPRB06}. Various vdW heterostructures have been experimentally realized via chemical vapor deposition (CVD), including multilayered and graphene-based configurations on transition-metal catalysts \cite{QiC18}. Therefore, the MnPS$_3|$WS$_2$ heterostructure is highly feasible for synthesis via similar CVD routes. To shed light on the inherent characteristics and potential applications of the MnPS$_3|$WS$_2$ heterojunction, we conduct a comprehensive investigation of its structural geometry, stability, and electronic structure. Our further analysis involves examining the effects of strain and external fields on the stability of the observed structures. In addition to providing insight into the potential of heterostructures like those for the development of spintronic and valleytronic devices of the next generation, this theoretical work offers essential understanding into the physics that lies beneath the surface for next-generation spintronic and valleytronic devices.

Altermagnetism, valley-contrasting physics, and Rashba spin splitting have each been pursued individually as routes to spintronic and valleytronic functionality, where altermagnets combine the stray-field-free compensated order of an antiferromagnet with ferromagnet-like spin splitting, valleytronics exploits the valley index as a non-charge information carrier, and Rashba coupling enables charge-spin interconversion through spin-momentum locking. Integrating all three functionalities into a single interface, rather than distributing them across different materials, is important in various respects. When these features coexist, they are not independent; the same symmetry-breaking mechanism that activates one typically constrains the others, and a single external stimulus can govern multiple responses at once. Therefore, we can simultaneously invert both the altermagnetic and valley-resolved spin splitting, establishing a single material as an integrated valleytronic-spintronic platform. Also, this coexistence is practically favorable for device integration: importing Rashba and valley-selective spin-orbit fields via proximity to separate layers introduces additional interfaces, each contributing lattice mismatch and disorder. A single interface that natively hosts all three effects removes the need for extra stacking and is largely independent in its tunability, offered by the electric field and strain. Therefore, MnPS$_3|$WS$_2$ interface offers two external handles that can push a single platform between qualitatively distinct functional regimes (e.g., type-I/type-II alignment, valley-degenerate/valley-split limits) without requiring a new material system.

The remainder of the article is organized as follows: In \cref{sec:method}, we describe the crystal structure, stacking configuration, and underlying symmetries of the MnPS$_3$ and WS$_2$ monolayers and their heterostructure. In \cref{sec:results}, we present the {\em ab initio} density functional theory (DFT) calculations for the monolayer as well as the heterostructure system, introduce the effective model Hamiltonians, and derive the corresponding expressions for the Berry curvature. We also investigate the effects of strain and external electric field on the heterostructure. Finally, \cref{sec:summary} summarizes our main results and conclusions.
\section{\label{sec:method}Crystal structure and Methodology}
\begin{figure}
\includegraphics[scale=0.265]{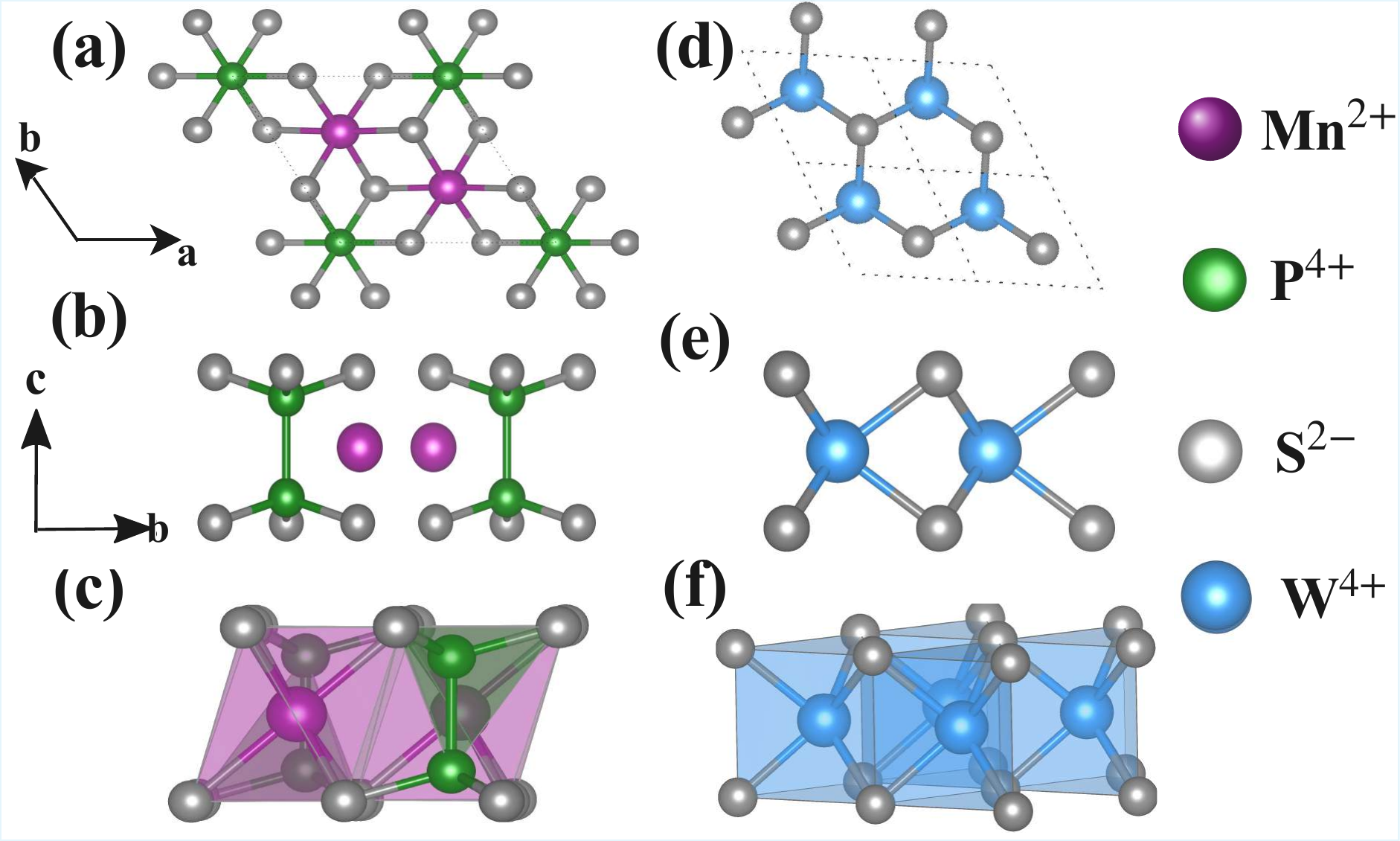}
\caption{\label{fig:Unit_cell}The unit cell structures of the individual monolayers and their corresponding crystal field environments for MnPS$_3$ and WS$_2$ are illustrated here. Panels (a) and (b) show the optimized top and side views of the MnPS$_3$ unit cell, respectively. MnPS$_3$ forms a layered honeycomb lattice in which Mn and S atoms constitute tilted octahedra, while the P$_2$ dimers form an inverted prism structure with surrounding S atoms, as depicted in (c). Similarly, (d) and (e) present the top and the side views of the WS$_2$ unit cell after structure optimization. WS$_2$ adopts a trigonal prismatic coordination, characteristic of the 2H-phase transition metal dichalcogenides, depicted in (f).}
\end{figure}
\begin{figure*}
\includegraphics[scale=0.26]{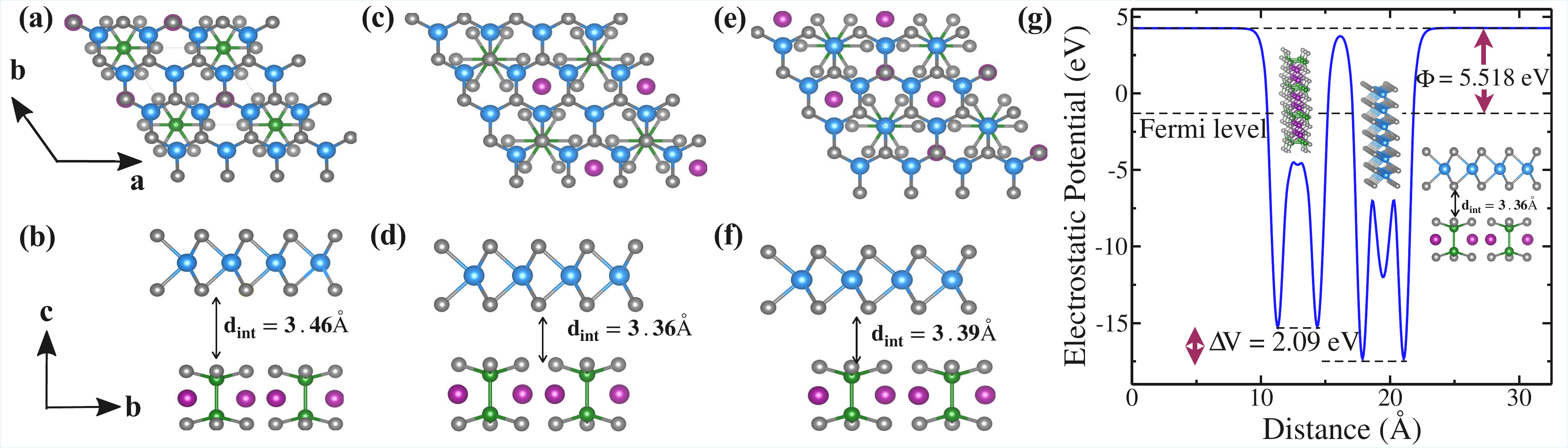}
\caption{\label{fig:Stacking}The top view and the side view of the simulated MnPS$_3|$WS$_2$ heterostructure with stacking configuration I are depicted in (a) and (b), respectively. Similarly, (c), (d), and (e), (f) show the top and side views of stacking configuration II and III, respectively. The variation of the electrostatic potential energy along the $z$ direction across the simulation cell is shown in (g), depicting the built-in potential difference in the heterostructure.}
\end{figure*}
Transition-metal phosphorus trichalcogenides are crystalline compounds with layered structures, characterized by weak vdW interactions between adjacent $M$P$X_3$ ($M$ = Mn, Fe, Ni; $X$ = S, Se) layers. The transition metal ions form a honeycomb lattice, each octahedrally coordinated by six sulfur atoms, while P$_2$ dimers are situated at the centers of the hexagons, resulting in a trigonal arrangement with S atoms. Monolayer MnPS$_3$ exhibits hexagonal symmetry in the $P\bar{3}_1m$ space group, mapping to the $D_{3d}$ point group \cite{NealPRB19}, with Mn, P, and S occupying the $2d$, $2e$, and $6k$ Wyckoff positions, respectively. Each Mn$^{2+}$ ion adopts a high-spin state with $S = 5/2$ due to the weak ligand field of the octahedral S$^{2-}$ anions. MnPS$_3$ belongs to the class of N\'eel-type antiferromagnets \cite{NiPRL21, ChuPRL20}. In this material, the honeycomb lattice naturally separates into two equivalent triangular sublattices, denoted by $A$ and $B$, whose spin moments satisfy $\vec{S}_{A}=-\vec{S}_{B}$, resulting in a finite N\'eel vector,
$$
\vec{L}=\vec{M}_{A}-\vec{M}_{B}\neq0,
$$
while the total magnetization remains exactly zero,

$$
\vec{M}=\vec{M}_{A}+\vec{M}_{B}=0.
$$
This compensated magnetic configuration is the defining characteristic of collinear antiferromagnetic order. The space group of monolayer MnPS$_3$, $P\bar{3}_1m$, features two-fold rotational symmetry along the $[1\bar{1}0]$, $[120]$, and $[210]$ axes, as well as mirror planes perpendicular to these directions. Additionally, there are three-fold anticlockwise and clockwise rotations about the $[001]$ axis, inversion, and the identity operation. MnPS$_3$ is a collinear antiferromagnet with two spin sublattices that possess equal but oppositely oriented magnetic moments. The combined $\mathcal{PT}$ symmetry operation is generally considered responsible for the energy band degeneracy characteristic of compensated magnetism. In the $P\bar{3}_1m$ group, local magnetic moments reverse under $\mathcal{PT}$, but a mirror operation about the $(010)$ plane, combined with $\mathcal{T}$, restores the original orientation, preserving a mirror-assisted $\mathcal{PT}$ symmetry in MnPS$_3$ monolayer. Hence, this symmetry operation relates both magnetic sublattices, ensuring a net zero magnetic moment. The 2H phase of WS$_2$ features two hexagonal planes of S atoms and a central hexagonal plane of W atoms, with W trigonally coordinated to the chalcogen layers in an ABA stacking sequence. The WS$_2$ monolayer exhibits hexagonal symmetry with space group $P\bar{6}m_2$. This space group includes two-fold rotational symmetry along the $[1\bar{1}0]$, $[120]$, and $[210]$ axes; mirror symmetry across the $[001]$, $[110]$, $[100]$, and $[010]$ planes; three-fold rotational symmetry both anticlockwise and clockwise about the $[001]$ axis; six-fold anticlockwise and clockwise rotational symmetry with its inversion about the $[001]$ axis; and the identity operation.

We employ density functional theory (DFT) using the {\scshape vasp} code \cite{vasp1, vasp2}, which utilizes a plane wave basis set along with the projector augmented wave (PAW) method \cite{paw}. The exchange-correlation potential is described using the generalized gradient approximation (GGA) due to Perdew-Burke-Ernzerhof (PBE) \cite{pbe}. In this study, we apply Grimme-D2 (DFT-D2) corrections within the GGA-PBE framework to address the impact of dispersive interactions on the electronic structure of heterogeneous bilayers \cite{GrimmeJCC04}, ensuring the inclusion of weak intra- and inter-layer bonding effects. A $\Gamma$-centered $\vec{k}$ mesh of $15 \times 15 \times 1$ has been employed for integration over the Brillouin zone, utilizing the tetrahedron method \cite{BlochlPRB94T}. To account for the strong onsite Coulomb repulsion arising from the localized Mn-$3d$ electrons, we adopted the DFT+$U$ approach \cite{DudarevPRB98} with an effective Hubbard parameter ($U_{\text{eff}}$) of $5$~eV. For self-consistent calculations that incorporate SOI, 3D band computations, and other property evaluations, we used a uniform $\vec{k}$-point mesh of $25 \times 25 \times 1$. An energy cutoff of 500~eV was set for the plane wave expansion of the electronic wave function. The atomic positions and lattice vectors have been optimized to minimize the force on each atom with a tolerance of $10^{-2}$~eV/\AA\ on the force. We used $10^{-5}$~eV per unit cell as the convergence criterion for self-consistent total energy in the DFT calculation.
\begin{figure*}
\includegraphics[scale=0.28]{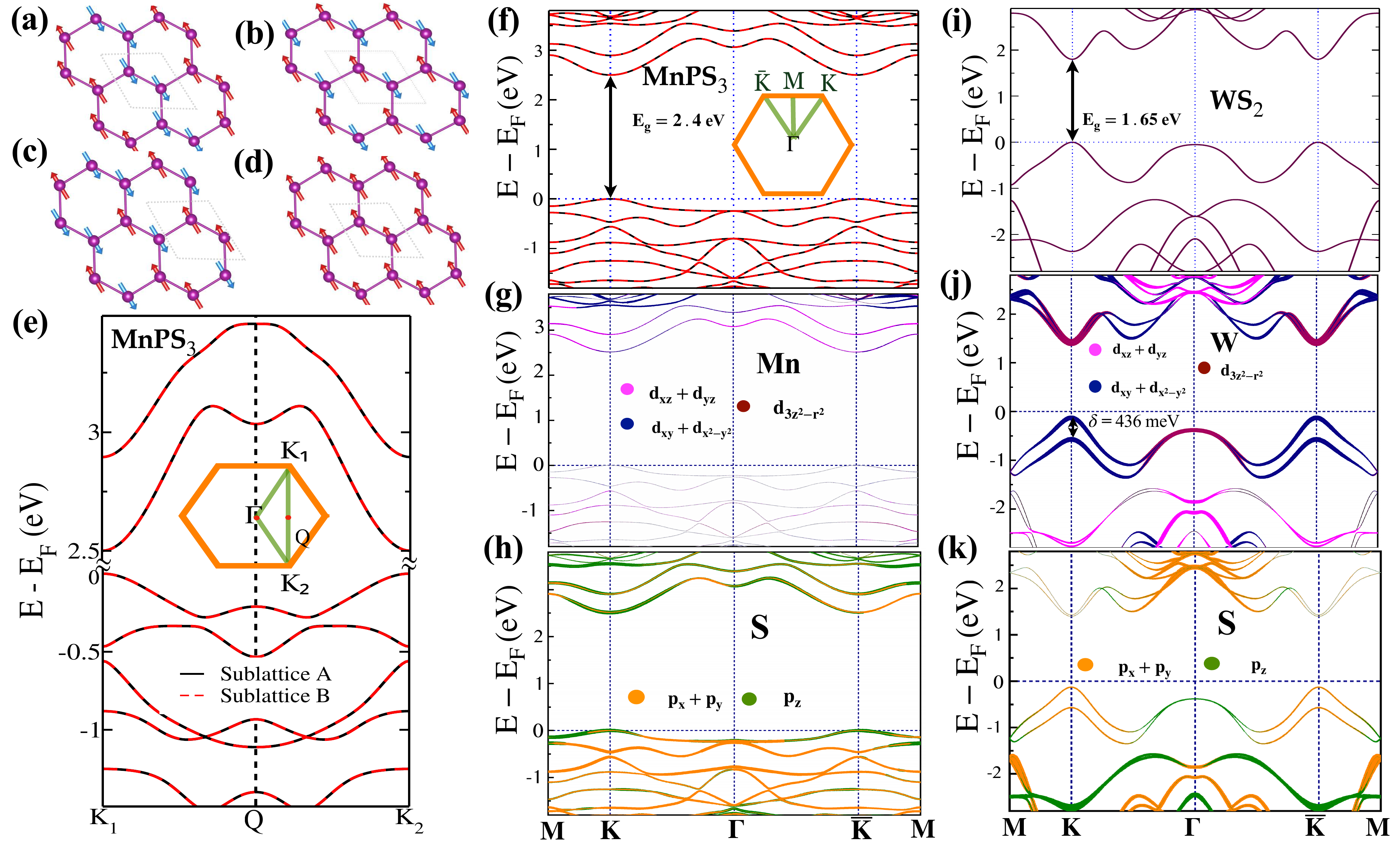}
\caption{\label{fig:Monolayers}The magnetic moment arrangements, spin-polarized, and orbital-projected band structures of MnPS$_3$ and WS$_2$ monolayers from our DFT calculations are presented here. Possible magnetic moment arrangements of MnPS$_3$ are illustrated in (a), (b), (c), and (d), termed as zigzag-type AFM, Néel-type AFM, stripy-type AFM, and ferromagnetic configurations, respectively. Panels (e) show the spin-polarized band dispersion of MnPS$_3$ along the $K_1 \to K_2$ path, with the inset showing the Brillouin zone, marking $\Gamma$, $K_1$, $K_2$, and $Q$-points. Panel (f) displays the spin-polarized band structure of MnPS$_3$ along the high-symmetry path $M \to K \to \Gamma \to \bar{K} \to M$, obtained from our DFT study. Panels (g) and (h) depict the orbital-projected band structures of MnPS$_3$, including SOI, highlighting the contributions from different atomic orbitals. For the WS$_2$ monolayer, panel (i) presents the spin-polarized band dispersion, while panels (j) and (k) show the corresponding orbital-projected band structures with SOI, emphasizing the orbital contributions to the electronic states.}
\end{figure*}
The in-plane lattice constants obtained from our calculations for MnPS$_3$ and WS$_2$ monolayers are $6.067$~\AA\ and $3.154$~\AA, respectively. The simulated unit cells and crystal field environments of MnPS$_3$ monolayer and WS$_2$ monolayer are illustrated in \cref{fig:Unit_cell}. \Cref{fig:Unit_cell}(a) and \cref{fig:Unit_cell}(b) depict the top and side-views, respectively, of the MnPS$_3$ unit cell, while the microscopic details of the structure are illustrated in \cref{fig:Unit_cell}(c). On the other hand, \cref{fig:Unit_cell}(d) and \cref{fig:Unit_cell}(e) depict the top and side views of the WS$_2$ unit cell, while \cref{fig:Unit_cell}(f) illustrates the trigonal prismatic coordination of the S atoms surrounding the W atoms in the 2H phase. We simulate the heterostructure of a unit cell of MnPS$_3$ monolayer with a $2\times2\times1$ supercell of WS$_2$, as depicted in \cref{fig:Stacking}, having the in-plane lattice constant of MnPS$_3$, $a = 6.067$~\AA, and the lattice vectors $\vec{a} = a \hat{x}$, $\vec{b} = -\frac{1}{2}a \hat{x} + \frac{\sqrt{3}}{2}a \hat{y}$, and $\vec{c} = c \hat{z}$. A vacuum of 28~\AA\ is introduced along the $z$-axis to prevent interactions between periodic images. We explored three distinct stacking configurations in the MnPS$_3|$WS$_2$ heterostructure, as illustrated in \cref{fig:Stacking}. In the first stacking configuration (see \cref{fig:Stacking}(a) (top view) and \cref{fig:Stacking}(b) (side view)), the WS$_2$ layer is placed directly above the MnPS$_3$ layer such that the W and S atoms fall alternately above the Mn sites of the Mn hexagon. In the second configuration (see \cref{fig:Stacking}(c) (top view) and \cref{fig:Stacking}(d) (side view)), the WS$_2$ layer is placed laterally so that the W atoms are aligned only with every alternate Mn site of the hexagon, leaving the remaining Mn positions unoccupied above. In the third configuration (see \cref{fig:Stacking}(e) (top view) and \cref{fig:Stacking}(f) (side view)), the WS$_2$ layer is displaced relative to the lower layer such that the S atoms now fall above alternate Mn atoms of the Mn hexagon, while the other Mn sites remain unoccupied from above. We refer to these stacking configurations as I, II, and III stacking. This approach allowed us to gain deeper insights into the electronic properties and magnetic behavior of the heterostructure beyond those of the individual components. Further analysis involves examining the effects of strain and external electric fields on the stability of the observed structures.
\section{\label{sec:results}Results and Discussions}
This section presents a detailed analysis of the electronic and magnetic properties derived from DFT calculations for monolayer MnPS$_3$, monolayer WS$_2$, and the MnPS$_3|$WS$_2$ vdW heterostructure. We explore these properties both in the absence and presence of SOI to understand its impact on the system. Additionally, our results include the effects of external factors, such as strain and applied electric field, on the stability and behavior of these materials.
\subsection{Electronic Structure}
We begin by identifying an optimum magnetic configuration for the system. Some of the probable magnetic configurations for MnPS$_3$ monolayer are depicted in \cref{fig:Monolayers}(a)-(d). Our calculations reveal an antiferromagnetic configuration of local magnetic moments where the nearest magnetic atom's moment points in the opposite direction, as illustrated in \cref{fig:Monolayers}(b), which leads to the lowest energy for the monolayer of MnPS$_3$, corresponding to $2/m.1^{\prime}$ magnetic point group. The anti-unitary rotation $2'_{010}$ transforms one magnetic sublattice into the other, opposite-spin magnetic sublattice. \Cref{fig:Monolayers}(e) and (f) display the band dispersions obtained from spin-polarized DFT calculations for the lowest-energy magnetic configuration of MnPS$_3$ monolayer along the paths $K_1 \to K_2$ (see the inset of \cref{fig:Monolayers}(e)) and $M\to K \to \Gamma \to \bar{K}\to M$ (see the inset of \cref{fig:Monolayers}(f)), respectively. The spin-polarized band structures of MnPS$_3$, shown in \cref{fig:Monolayers}(e) and \cref{fig:Monolayers}(f), reveal that the opposite-spin bands are perfectly degenerate owing to the $\mathcal{PT}$-symmetry, forming two energy-degenerate valleys at the $K$ and $\bar{K}$ points. The band dispersion in panel (f) reveals a direct band gap of 2.4~eV at the $K$ point, with a projected magnetic moment of $4.59\, \mu_B$ at each Mn. Besides underestimating the band gap, DFT calculations within a non-local plane-wave basis set also slightly underestimates the projected magnetic moments compared to experiments \cite{BrecIC79}. The orbital-projected band dispersions with SOI are displayed in \cref{fig:Monolayers}(g) and \cref{fig:Monolayers}(h), showing a negligible effect of SOI on the band dispersion, corresponds to the hybridized S $p$ and Mn $d$ orbitals (see \cref{fig:Monolayers}(g)), while VBM primarily arises from the S $p$ orbitals (see \cref{fig:Monolayers}(h)).
The spin-polarized band structure of WS$_2$ monolayer reveals similar features at the $K$ and $\bar{K}$ points, with a direct band gap of 1.53~eV, as shown in \cref{fig:Monolayers}(i). However, upon considering the SOI, the band dispersion exhibits well-separated valleys \cite{LiPNAS13}, establishing a large spin splitting (0.426 to 0.433~eV) at the $K$ point in the VBM. We observe that the VBM has two local maxima at the $K$ and $\Gamma$ points, while the CBM has two local minima, one at the $K$ point and the other between the $\Gamma$ and $K$ points, as shown in \cref{fig:Monolayers}. In the VBM, the local maxima at the $K$ points are primarily influenced by W $d_{x^2 - y^2}$ and $d_{xy}$ bonding states. In contrast, the local maxima at the $\Gamma$ point are predominantly derived from W $d_{3z^2-r^2}$ states. In the CBM, the local minima at the $K$ points are mainly composed of W $d_{3z^2-r^2}$ states.
\begin{figure*}
\includegraphics[scale=0.29]{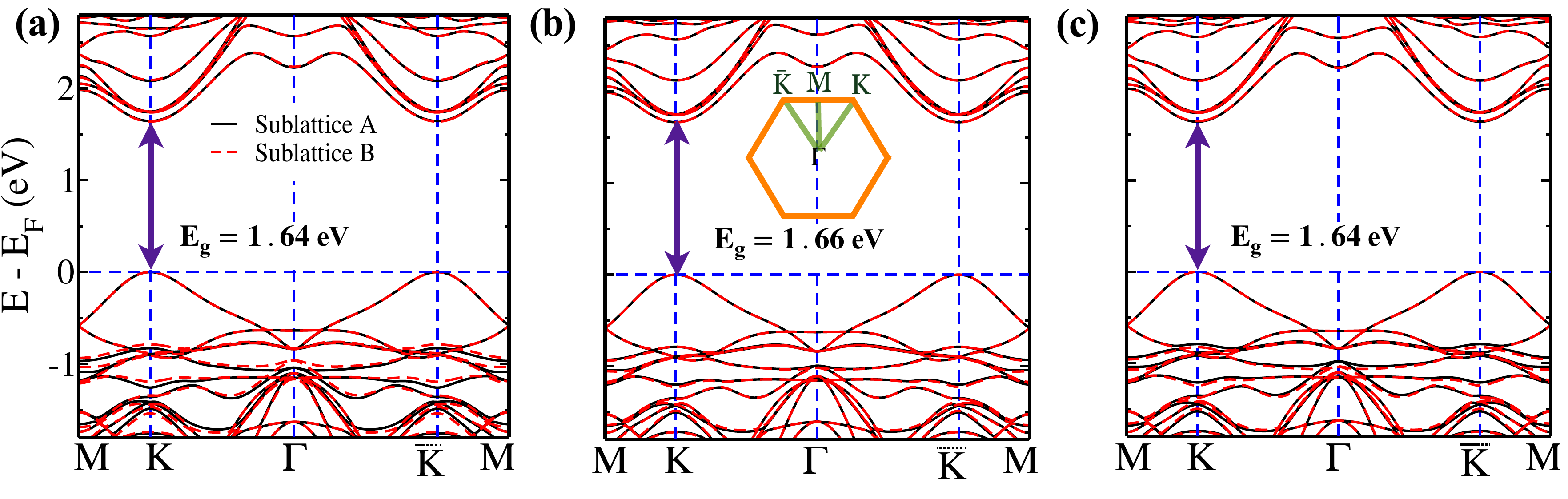}
\caption{\label{fig:spin_pol} Panels (a), (b), and (c) show the spin-polarized band dispersion of the heterostructure, for the stacking configurations I, II, and III, respectively, along the high-symmetry direction $M \to K \to \Gamma \to \bar{K} \to M$ from our DFT calculation. The band dispersion for three cases indicates a direct band gap ranging from $1.64$ to $1.66$~eV at the $K$ and $\bar{K}$ points. The inset of (b) shows the two-dimensional hexagonal Brillouin zone, marking the relevant high-symmetry points.}
\end{figure*}
\begin{figure*}
\includegraphics[scale=0.28]{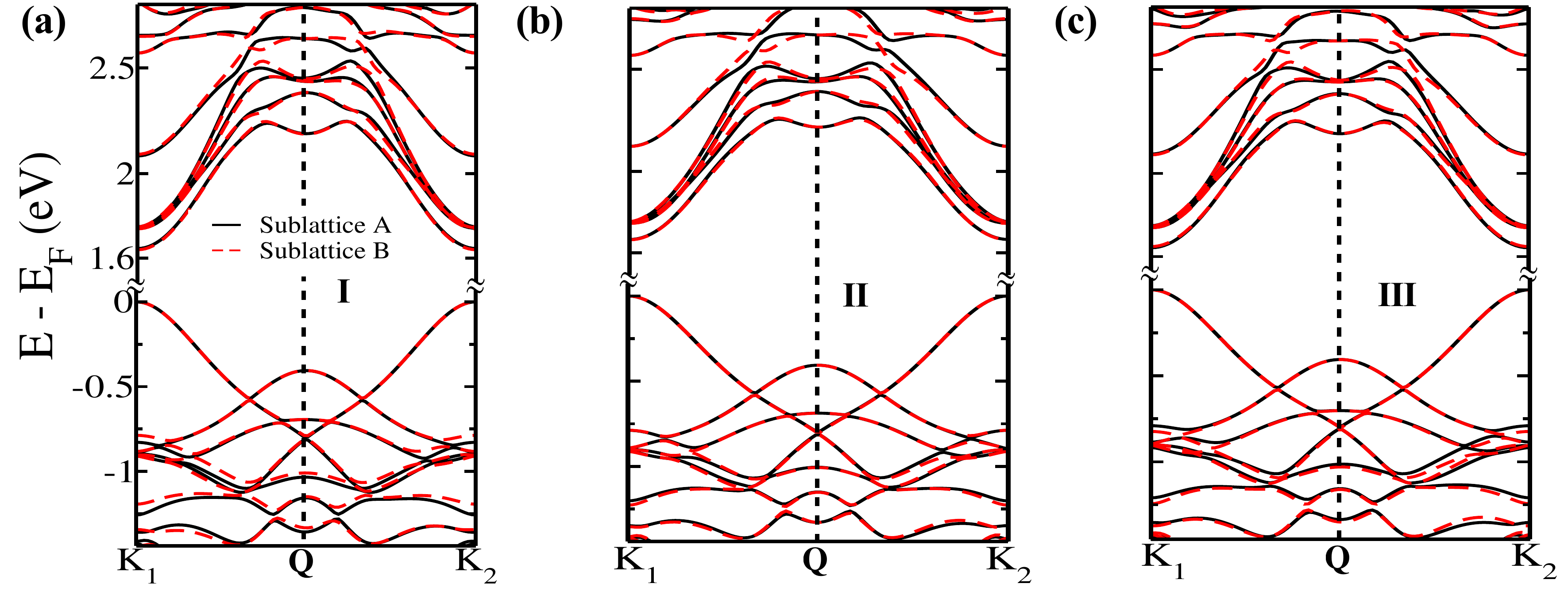}
\caption{\label{fig:altermagnet} The band dispersion with broken degeneracy along the $K_1 \to Q \to K_2$ path, as marked in the inset of \cref{fig:Monolayers}(c), is shown here. Panels (a), (b), and (c) show the spin-polarized band dispersion of MnPS$_3|$WS$_2$ heterostructure for the stacking configurations I, II, and III, respectively, as obtained from our DFT calculations. These bands show that the opposite-spin bands from different sublattices are not degenerate, except at the $Q$-point.}
\end{figure*}
To estimate the stability of the three stacking configurations of the MnPS$_3|$WS$_2$ heterostructure considered here, we calculate the corresponding binding energies, $E_b$, defined as
\begin{equation}
    E_b = E_{\text{MnPS}_3|\text{WS}_2} - E_{\text{MnPS}_3} - E_{\text{WS}_2}, \label{eq:BindingEnergy}
\end{equation}
where $E_{\text{MnPS}_3|\text{WS}_2}$, $E_{\text{MnPS}_3}$, and $E_{\text{WS}_2}$ are the total energies of the heterostructure, the MnPS$_3$ monolayer, and the WS$_2$ monolayer, respectively, obtained from the DFT calculations. All the configurations exhibit negative binding energies, indicating energetic stability, and share similar features in their band structures. The heterogeneous stacking of layers breaks the out-of-plane structural symmetry, leading to an internal electric field across the layers. The average electrostatic potential of the MnPS$_3|$WS$_2$ heterostructure is illustrated in \cref{fig:Stacking}(d). The WS$_2$ layer has a deeper potential than that of the MnPS$_3$ layer, resulting in the formation of a built-in electric field across the interface from the MnPS$_3$ layer to the WS$_2$ layer. The calculated potential drop is 2.06 eV, which considerably impacts the electronic and magnetic properties at the interface. As schematically illustrated magnetic configurations in \cref{fig:Monolayers}(a)-(d), we conduct DFT calculations for these potential magnetic configurations, including Néel-type AFM, zigzag-type AFM, stripy-type, and ferromagnetic configurations, to determine the energetically favorable ground state for the heterostructure. On comparing the energies from the self-consistent calculation, the N\'eel-type AFM configuration, with zero net magnetic moment, is the lowest energy magnetic configuration. The spin-polarized band dispersions for stacking configurations I, II, and III are shown in \cref{fig:spin_pol}(a), \ref{fig:spin_pol}(b), and \ref{fig:spin_pol}(c), respectively, suggesting a direct band gap of $\sim$1.65~eV and type-II band alignment at the $K$ and $\bar{K}$ points, with a perfect degeneracy of both sublattice bands along the high symmetry direction $M\to K \to \Gamma \to \bar{K}\to M$ for stacking configurations. Several bands from the opposite magnetic sublattices show broken degeneracy near $-1$~eV and above 2~eV across all stacking configurations, owing to dissimilar local environments. The two layers produce a weak distortion in the sublattices, and the local electrostatic environment experienced by each Mn atom depends on the specific interfacial atomic arrangement, whether a W or an S atom of the neighboring layer lies closest across the van der Waals gap. This inequivalence of the local atomic environments lifts the degeneracy of the corresponding Mn-derived bands, with the splitting most pronounced in the energy windows where these bands retain significant Mn character.
\subsection{Spin splitting without SOI}
\begin{figure}
\includegraphics[scale=0.215]{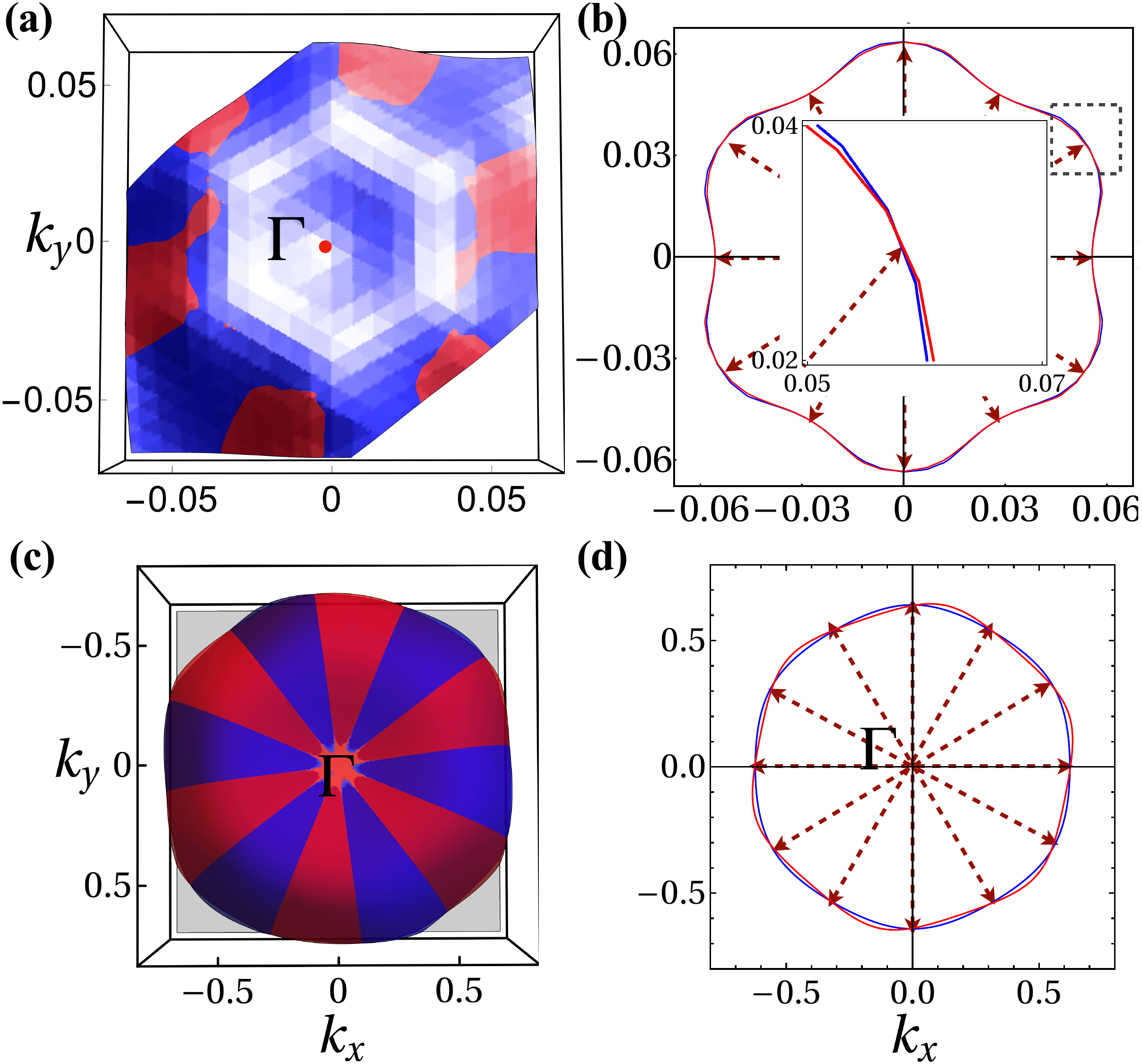}
\caption{\label{fig:alt_model}Panels (a) and (b) show the spin-polarized 3D band dispersion of MnPS$_3|$WS$_2$ heterostructure and the isoenergetic contours at $E - E_F = 2.2$~eV, respectively, as obtained from our DFT calculations. The band dispersion along the high-symmetry path of the hexagonal Brillouin zone reveals lifted degeneracy except at the nodal points, as highlighted in the inset of (b). Panels (c) and (d) show the 3D band dispersion $\varepsilon^{\pm}(k_x, k_y)$ and the corresponding isoenergetic contours, respectively, as obtained by solving the model Hamiltonian (see \cref{eq:Q-point Model,eq:Eigenvalue}). The arrows mark the spin-degenerate lines for $i$-wave altermagnet. Inset of (b)shows a close-up of the region marked by the square, showing the spin splitting of the bands.}
\end{figure}
The band dispersions along the high symmetry direction $K_1\to K_2$, displayed in \cref{fig:altermagnet}(a), \ref{fig:altermagnet}(b), and \ref{fig:altermagnet}(c) for stacking configurations I, II, and III, respectively, exhibit broken degeneracy through the bands due to asymmetry condition, unlike the MnPS$_3$ monolayer, where Kramers' degeneracy in the band dispersion is preserved owing to the presence of multiple symmetry operations that map one spin sublattice onto the other. In contrast, MnPS$_3$ hosts a non-relativistic splitting for the MnPS$_3|$WS$_2$ heterostructure, due to the presence of three-fold $3_{001}$ rotational symmetry operation in the non-magnetic point group. While the anti-unitary two-fold rotation operations $2^{\prime}_{010}$ connect the opposite magnetic sublattices, with missing inversion symmetry. Besides breaking the inversion symmetry, stacking of WS$_2$ on top of MnPS$_3$ generates the interlayer anisotropy of the electronic crystal potentials for opposite spin magnetic sublattices as W and S atoms occupy inequivalent in-plane lattice positions. This interlayer-induced anisotropy removes the degeneracy protection present in the monolayer and enables altermagnetic splitting at $Q$-points (see \cref{fig:Monolayers}(e) (inset)) \cite{AbsorPRB16}. The MnPS$_3|$WS$_2$ stacking reduces the symmetry to the point group $C_3$ and the magnetic little co-group $2^{\prime}$ around $Q$-point that has been identified as capable of hosting altermagnetic spin splitting in magnetic systems \cite{CheongNPJQM25}. The band dispersion obtained from our DFT calculation show the broken degeneracies along the $K_1 \to K_2$ high-symmetry path shown in \cref{fig:alt_model}(a), the two spin channels are non-degenerate almost everywhere in the Brillouin zone, except at a discrete set of nodal lines where the bands touch; the corresponding isoenergetic contours in \cref{fig:alt_model}(b) show an alternating spin-split pattern with six nodal lines, a signature of $i$-wave altermagnetism.
\subsubsection{Effective two-band model Hamiltonian at the $Q$-point without SOI}
In the MnPS$_3|$WS$_2$ heterostructure, the underlying magnetic symmetry keeps the bands spin-degenerate along certain directions in the Brillouin zone. However, once we move away from the $K$-point, a small but clear altermagnetic spin splitting appears along the path $K_1 \to K_2$ (see \cref{fig:Monolayers}(e) (inset)). To understand this better, we develop a symmetry-guided two-band effective Hamiltonian constrained by the magnetic point group $2^{\prime}$, so that the resulting model captures the symmetry-allowed terms. Under these symmetries, the Hamiltonian around the nodal point $Q$ in momentum space takes the form
\begin{equation}
   H = H_0 + \left[ \lambda \left( k_+^6 + k_-^6 \right) + \frac{\mu}{2i} \left( k_+^6 - k_-^6 \right) \right] \sigma_z
\label{eq:Q-point Model}
\end{equation}
where $k_{\pm} = k_x \pm ik_y$.
The first term of the Hamiltonian, $H_0$, leads to the energy $E_0 = [t(k_x^2+k_y^2) + \eta (k_x^2+k_y^2)^2]$, representing an isotropic dispersion, with the quartic correction capturing deviations from simple parabolic behavior. The first part of the second term preserves the magnetic point-group symmetry and introduces a six-fold in-plane angular modulation of the dispersion, identical to $2k^6\cos(6\theta)$, consistent with the allowed little co-group $2^\prime$ at the $Q$-point. The second part of the second term contributes to the symmetry-allowed spin splitting by generating an antisymmetric sixth-order angular dependence, which is identical to $2ik^6 \sin(6 \theta)$ in the polar coordinates, explicitly capturing the odd angular harmonic permitted by crystal symmetry. These sixth-order harmonics naturally generate the characteristic sign-alternating structure of the spin splitting in the momentum space that defines the $i$-wave altermagnetism. Expanding the basis function in terms of $k_x$ and $k_y$, the spin-dependent part of the Hamiltonian, including all the relevant sixth-order symmetry allowed contributions, the energy eigenvalues take the form
\begin{align}
E_{\pm}(k_x,k_y) =& E_0 \pm \big[ 2\lambda \left( k_x^6 - k_y^6 + 15 k_x^2 k_y^4 - 15 k_x^4 k_y^2 \right) \nonumber \\
&+ \mu \left( 6 k_x^5 k_y - 20 k_x^3 k_y^3 + 6 k_x k_y^5 \right) \big]
\label{eq:Eigenvalue}
\end{align}
The dispersion obtained from the symmetry-adapted effective Hamiltonian reproduces the spin-splitting features around the $Q$-point, as obtained from our DFT results (see \cref{fig:alt_model}). \cref{fig:alt_model}(c) and \cref{fig:alt_model}(d) show the analogous quantities $E_\pm (k_x, k_y)$ obtained from the model, reproducing both the nodal-line positions and the splitting pattern found in the DFT results. This direct correspondence between the band dispersions obtained from DFT calculations and the model indicates that the symmetry-allowed sixth-order terms in the effective Hamiltonian capture the essential features of the altermagnetism.
\subsection{Spin splitting with SOI}
\begin{figure*}
\includegraphics[scale=0.267]{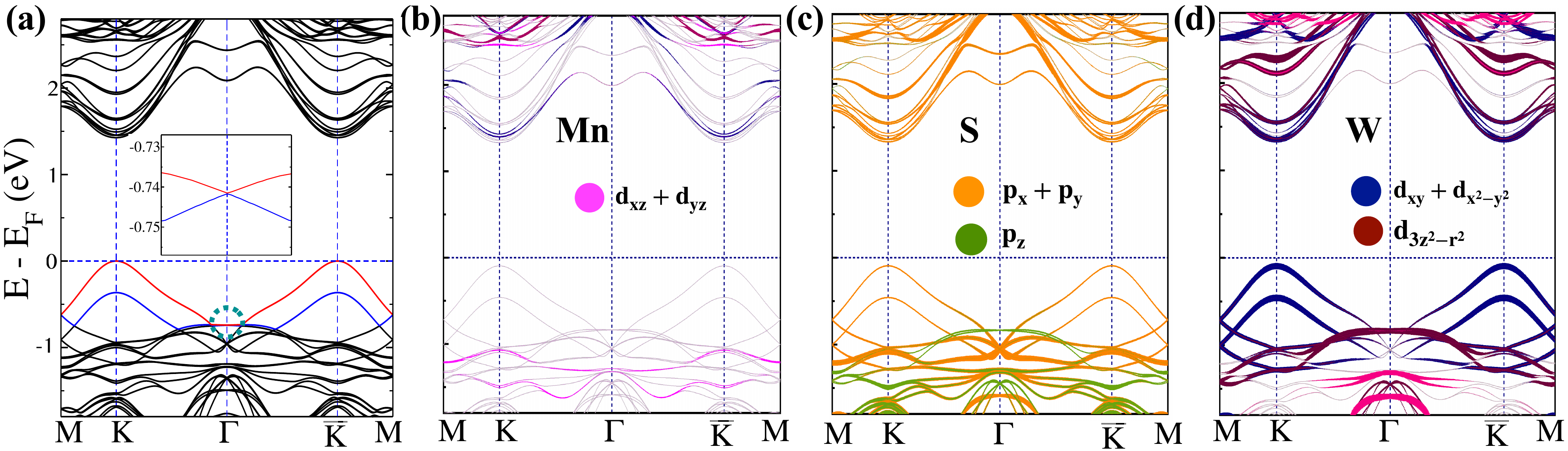}
\caption{\label{fig:Rashba}The spin-polarized band dispersion with the SOI and the orbital projected bands obtained from the DFT results are shown here. Panel (a) depicts band dispersion including SOI and the presence of the Rashba-like splitting in VBM. Panels (b), (c), and (d) show the orbital projected bands of Mn, S, and W, respectively. The inset of (a) highlights the Rashba-like splitting in the marked bands near the VBM.}
\end{figure*}
All the stacking configurations show similar projected magnetic moments of $\sim 4.61~\mu_B$ at each Mn atom, which is typically slightly underestimated within a plane-wave basis set. Our DFT calculations using $U_\text{eff} = 5$~eV, including SOI, reveal a negligible difference in magnetic anisotropy energy across in-plane and out-of-plane directions.

\begin{figure}
\includegraphics[scale=0.225]{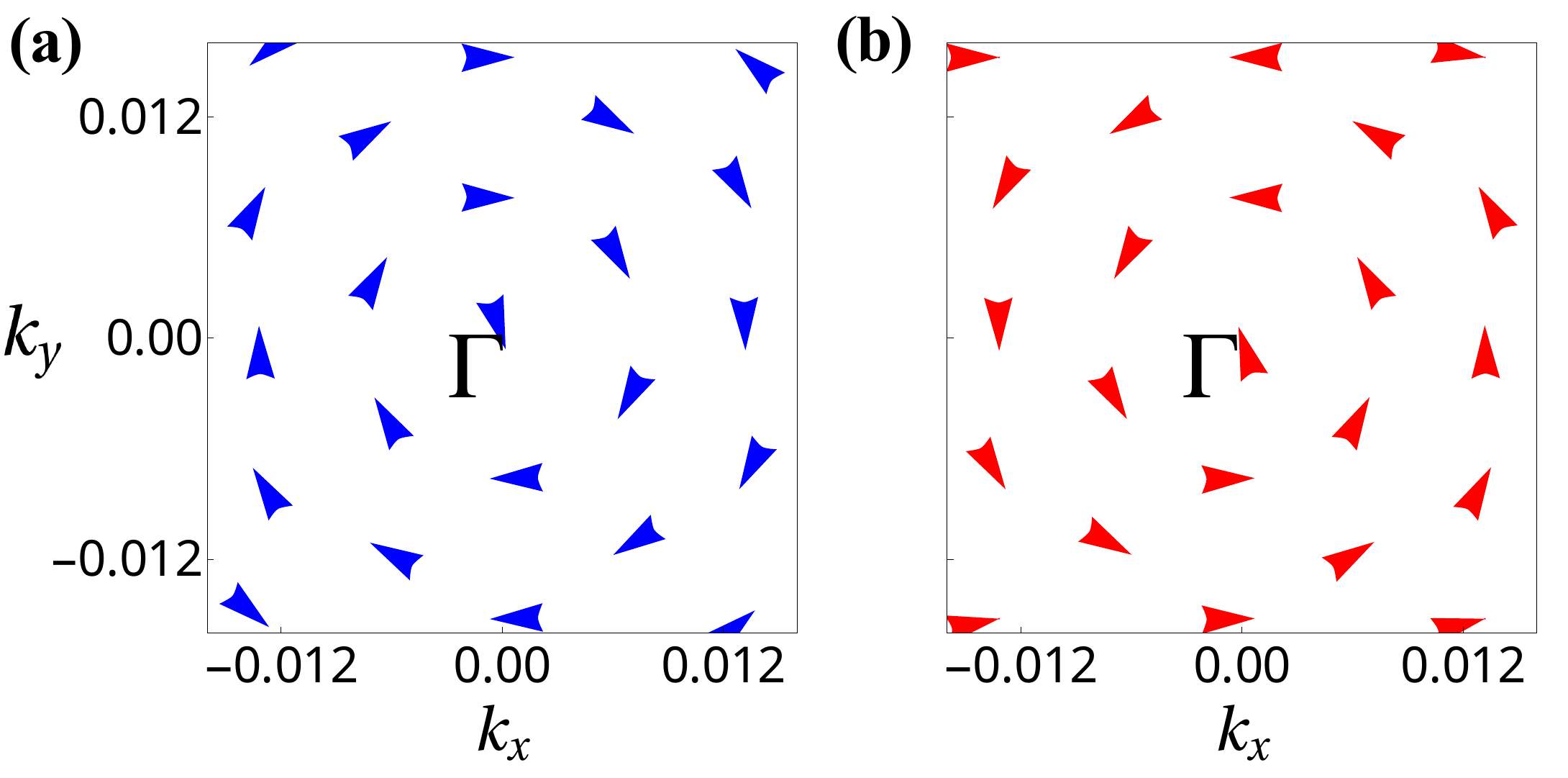}
\caption{\label{fig:Rashba_model}The spin-textures for the highest pair of valence bands at the $\Gamma$-point obtained from the DFT calculations, as marked in \cref{fig:Rashba}(a), are displayed in (a) and (b).}
\end{figure}

\begin{figure*}
\includegraphics[scale=0.29]{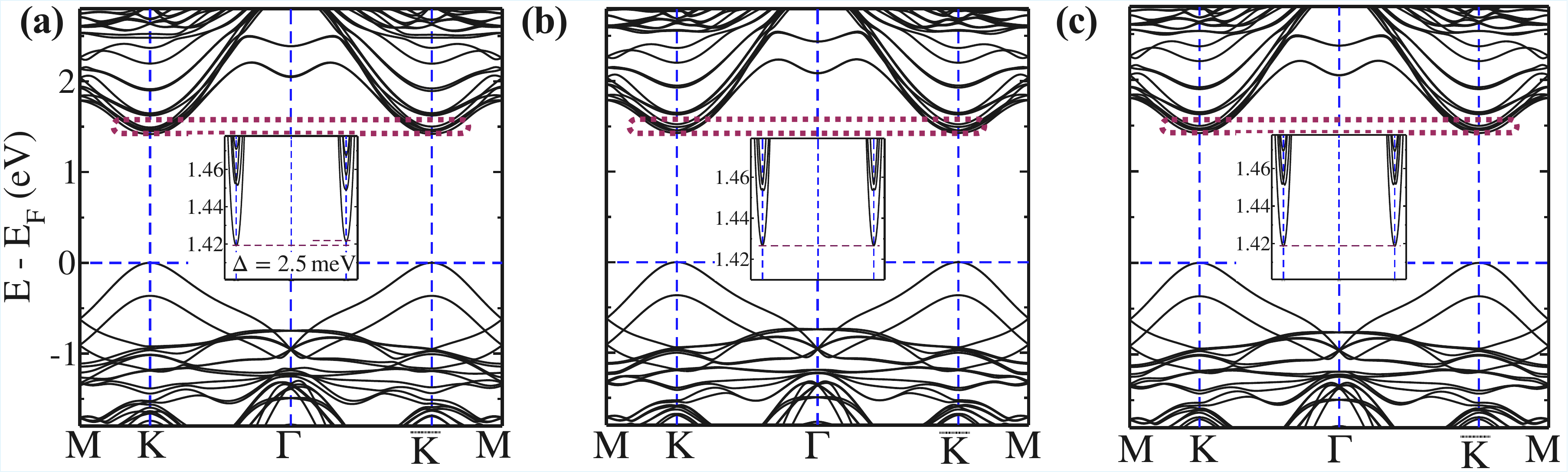}
\caption{\label{fig:with_SOC}The band dispersion of MnPS$_3|$WS$_2$ heterostructure with SOI and the valley splitting is presented here. Panels (a), (b), and (c) show the band dispersion of the heterostructure with SOI, for the stacking configurations I, II, and III, respectively, along the high-symmetry direction $M\to K \to \Gamma \to \bar{K}\to M$ from our DFT calculation. The inset images in panel (a), (b), and (c) show the close view  of the CBM near the $K$ and $\bar{K}$ point, demonstrating a non-zero valley-splitting for configuration I, while no valley-splitting for the cases of configurations II and III.}
\end{figure*}

The band dispersions due to SOI show significant splitting in the valence bands compared to the conduction bands across the three configurations. Both the $K$ and $\bar{K}$ valleys in the VBM exhibit considerable valley out-of-plane spin splitting; more importantly, a Rashba-like spin splitting in the bands is also observed in the $\vec{k}$-space at the $\Gamma$ point, as illustrated in \cref{fig:Rashba}(a) and highlighted in the inset. The projected bands in \cref{fig:Rashba}(b), \ref{fig:Rashba}(c), and \ref{fig:Rashba}(d) suggest that splitting in VBM at the $\Gamma$ point primarily corresponds to hybridization of W $d_{3z^2-r^2}$ and S $p_z$ orbitals. 
To analyze the following pair of bands, we plot energy $E$ as a function of $k_x$ and $k_y$ (3D bands) and corresponding spin textures. The projected spin vectors $S_x(\vec{k}) \hat{x} + S_y(\vec{k}) \hat{y}$ around the $\Gamma$ point form helical patterns (see \cref{fig:Rashba_model}(a) and \ref{fig:Rashba_model}(b)). This configuration reflects the characteristic spin textures associated with a Rashba-type spin-orbit interaction \cite{FrantzeskakisPRB11, GanguliPRB25, KumarPRB22, ChakrabortyPRB20}. In the CBM of MnPS$_3|$WS$_2$ band dispersion, a distinct valley splitting of 2.5~meV emerges in the stacking configuration I (see \cref{fig:with_SOC}(a)), whereas no appreciable valley splitting is observed for the other two stacking geometries (see \cref{fig:with_SOC}(b) and \ref{fig:with_SOC}(c)). This selective splitting originates from the substantially stronger crystalline sublattice anisotropy in the stacking configuration I, where W or S atoms of the WS$_2$ layer are positioned alternately above the Mn atoms in MnPS$_3$. Such an environment enhances Mn-W and Mn-S orbital overlap, particularly between Mn $d$-orbitals and the W $d$-dominated conduction bands of MnPS$_3|$WS$_2$. The resulting interlayer coupling breaks the residual valley degeneracy at the $K$ and $\bar{K}$ points. In contrast, in the remaining stacking configurations, the relative placement of W and S with respect to Mn atoms leads to significantly weaker orbital hybridization. The reduced overlap suppresses interlayer-induced perturbations to the local crystal field and thus fails to lift the valley degeneracy \cite{WangNPJCM22}. Consequently, the absence of strong Mn-W or Mn-S coupling in these geometries prevents the appearance of valley splitting in the CBM.
\subsubsection{Effective two-band model Hamiltonian at $\Gamma$ point with SOI}
\begin{table}
\caption{\label{tab:symmetry} Transformation of momentum $\{k_x,k_y\}$ and Pauli matrices $\{\sigma_x,\sigma_y\}$ at the $\Gamma$ point under the effect of $\{3^+_{001}|0\}$ and $\{3^-_{001}|0\}$ operations are shown here.}
\begin{ruledtabular}
\begin{tabular}{l c c}
Symmetry operation & $\{k_x, k_y\}$ & $\{\sigma_x, \sigma_y\}$ \\
\hline
$3^{+}_{001} \equiv e^{-i\frac{\pi}{3}\sigma_z}$ 
&
$\begin{pmatrix}
-\tfrac{1}{2}k_x + \tfrac{\sqrt{3}}{2}k_y \\
-\tfrac{\sqrt{3}}{2}k_x - \tfrac{1}{2}k_y
\end{pmatrix}$
&
$\begin{pmatrix}
-\tfrac{1}{2}\sigma_x + \tfrac{\sqrt{3}}{2}\sigma_y \\
-\tfrac{\sqrt{3}}{2}\sigma_x - \tfrac{1}{2}\sigma_y
\end{pmatrix}$
\\
$3^{-}_{001} \equiv e^{i\frac{\pi}{3}\sigma_z}$ 
&
$\begin{pmatrix}
-\tfrac{1}{2}k_x - \tfrac{\sqrt{3}}{2}k_y \\
\tfrac{\sqrt{3}}{2}k_x - \tfrac{1}{2}k_y
\end{pmatrix}$
&
$\begin{pmatrix}
-\tfrac{1}{2}\sigma_x - \tfrac{\sqrt{3}}{2}\sigma_y \\
\tfrac{\sqrt{3}}{2}\sigma_x - \tfrac{1}{2}\sigma_y
\end{pmatrix}$
\\
\end{tabular}
\end{ruledtabular}
\end{table}
The MnPS$_3|$WS$_2$ heterostructure exhibits $C_{3}$ point group symmetry, lacking an inversion center. This absence of inversion symmetry and SOI associated with W atoms, combined with a built-in electrostatic potential difference across the heterostructure, leads to momentum-dependent Rashba spin splitting. Our DFT calculations reveal that this Rashba-like spin splitting is prominent around the $\Gamma$ point, as illustrated in \cref{fig:Rashba_model}(a) and (b). We find $C_{3}$ little-group symmetry at the $\Gamma$ point, which allows both linear and cubic terms in $\vec{k}$ in the effective Hamiltonian. With the help of a symmetry-allowed low-energy $\vec{k} \cdot \vec{p}$ effective Hamiltonian containing only linear Rashba terms \cite{GanguliPRB25},
\begin{equation}
    H = H_0 + \alpha_R(k_x \sigma_y - k_y \sigma_x),
    \label{eq:GammaPointModel}
\end{equation}
where $H_0$ denotes the free-electron Hamiltonian, we  determine the Rashba coefficient $\alpha_R \approx 0.4$ by fitting the DFT-obtained bands to \cref{eq:GammaPointModel}.
\subsubsection{$\vec{k}\cdot \vec{p}$ model Hamiltonian at $K/\bar{K}$ point with SOI}
\begin{figure}
\centering
\includegraphics[scale=0.25]{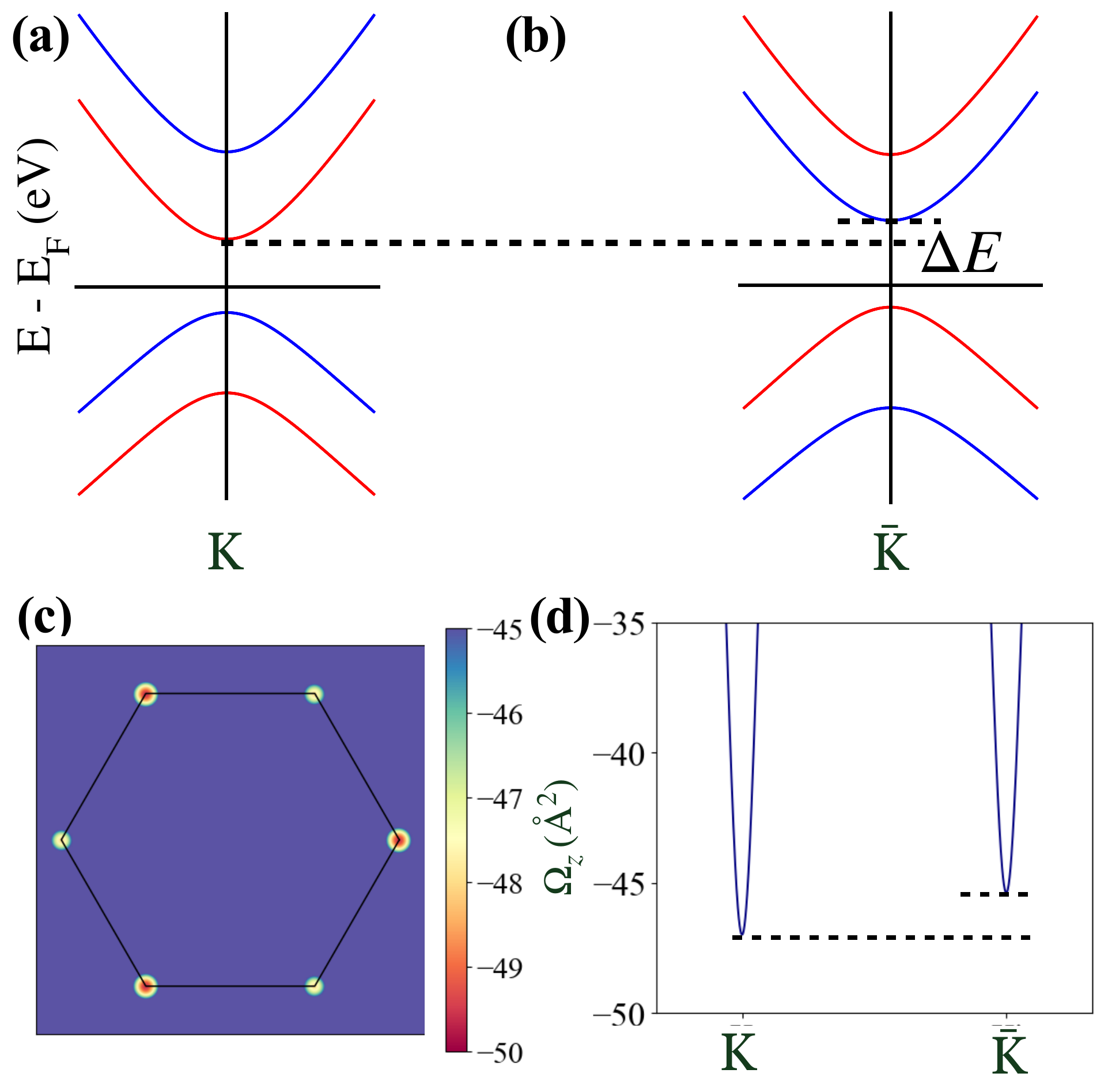}
\caption{\label{fig:inequivalent_model}The band splitting at the high-symmetry points $K$ and $\bar{K}$, along with the corresponding Berry curvature obtained from the model Hamiltonian, is presented here. Panels (a) and (b) display the band dispersion at the $K$ and $\bar{K}$ points, respectively, from the eigenvalues obtained from equation $5$. Panels (c) and (d) show the Berry curvature of the corresponding bands, computed using the fitted parameters of the model Hamiltonian: $v_F = 1.23 \times 10^5$~m/s, $\Delta = -0.075\,\mathrm{eV}$, $\delta = 0.07\,\mathrm{meV}$, $\lambda_i = 0.072\,\mathrm{meV}$, $\lambda_u = 0.06\,\mathrm{meV}$, and $\lambda_v = 0.072\,\mathrm{meV}$. The color gradient represents the magnitude of the $z$-component of the Berry curvature.}
\end{figure}
Spin-valley locking and valley polarization are some of the most useful features of 2D materials and heterostructures, with a variety of interesting outcomes and envisaged applications. Since our DFT calculations show negligible MAE, an out-of-plane easy-axis orientation is adopted for the investigation of valley features present in the band dispersion. In an attempt to explain our findings from DFT calculations (see \cref{fig:with_SOC}), here we develop an effective $\vec{k} \cdot \vec{p}$ model Hamiltonian to elucidate the microscopic origin of the spin-valley locking and the associated valley polarization at the high-symmetry points $K$ and $\bar{K}$ of the hexagonal Brillouin zone. At these points, the electronic states transform according to the little group $C_3$, imposing \{$3^+_{001}|0$\} and \{$3^-_{001}|0$\} rotational symmetry in the presence of SOI, forming the effective Hamiltonian \cite{XuAPL23}
\begin{equation}
H_{K/\bar{K}} = H_0 + H_{\text{Dirac}} + H_{\text{mass}} + H_{\text{SOI}},
\label{eq:kpValley}
\end{equation}
with
\begin{align}
    H_{\text{Dirac}} &= v_F \left( s_0 \tau_z \sigma_x k_x + s_0 \tau_0 \sigma_y k_y \right), \nonumber \\
    H_{\text{mass}} &= \Delta s_0\tau_0\sigma_z + \delta s_z\tau_0\sigma_z, \text{ and} \nonumber \\
    H_{\text{SOI}} &= \left( \lambda_i s_z + \lambda_u s_0 \right) \tau_z \sigma_z + \lambda_v s_z \tau_z \sigma_0. \nonumber
\end{align}
In this framework, $v_F$ denotes the Fermi velocity and $s_i$, $\tau_i$, and $\sigma_i$ $(i=x,y,z,0)$ are the Pauli matrices acting on the spin, valley, and sublattice degrees of freedom, respectively, with $i=0$ denoting the $2 \times2$ identity matrix. The first term $H_0$ describes the isotropic, free-electron kinetic dispersion common to both valleys and spin states. The second term $H_{\text{Dirac}}$ represents the massless Dirac Hamiltonian, which describes the low-energy electronic states near the $K$ and $\bar{K}$ valleys and gives rise to the characteristic conical band dispersion. The inequivalent valleys are distinguished by indices $\pm$. $H_{\text{mass}}$ consists of the symmetry-allowed sublattice mass terms that couple to $\sigma_z$. The spin-independent mass $\Delta$ breaks the equivalence between the two sublattices and opens a gap at the $K$ and $\bar{K}$ points, while the spin-dependent mass parameter $\delta$ lifts the spin degeneracy by generating spin-dependent band gaps. Combined with the spin-orbit interaction, these mass terms contribute to the effective Dirac mass, which controls the Berry curvature and the resulting valley-dependent electronic properties. The final term $H_{\text{SOI}}$ captures the intrinsic spin-valley coupling and valley polarization in the conduction band \cite{CaiPRB13}. It comprises three physically distinct contributions: (i) an intrinsic spin-orbit term $\lambda_i s_z \tau_z \sigma_z$, which is odd under both time reversal and spatial inversion and produces spin splitting that changes sign between $K$ and $\bar{K}$, (ii) sublattice term $\lambda_u s_0 \tau_z \sigma_z$, which is spin-independent but valley-antisymmetric, renormalizing the sublattice gap differently in the two valleys, and (iii) $\lambda_v s_z \tau_z \sigma_0$, valley-Zeeman term which acts uniformly across sublattices and generates a net spin splitting that is locked to the valley index, directly producing the observed spin-valley splitting, shown in \cref{fig:inequivalent_model}(a) and \ref{fig:inequivalent_model}(b).

\begin{figure*}
\includegraphics[scale=0.26]{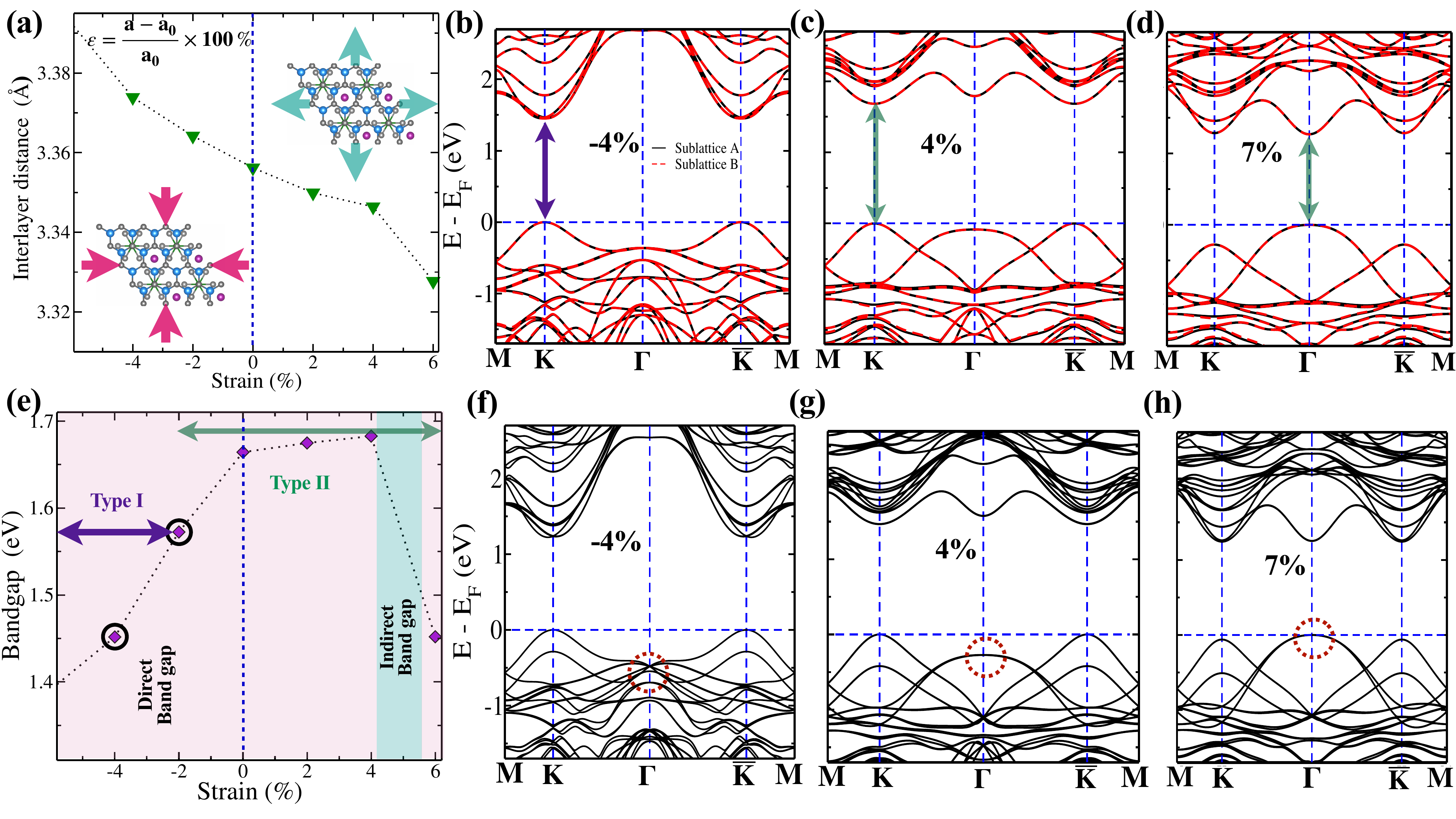}
\caption{\label{fig:strain} In-plane biaxial strains are applied on the MnPS$_3|$WS$_2$ heterostructure, which affect the lattice parameters equally along crystallographic $a$ and $b$ axes are shown schematically, and the variation of distance between each layer as a function of strain is shown here. Panel (a) shows the schematic for applying an in-plane tensile and compressive strain in our type-II stacking configuration of the heterostructure, respectively, along with the variation of interlayer distance corresponding to tensile (positive values) and compressive strain (negative values). Panels (b), (c), and (d) show the band dispersion without SOI, while panels (f), (g), and (h) show the band dispersion with SOI, under $-4\%$ (compressive) strain, $4\%$ (tensile) strain, and $7\%$ (tensile) strain,  respectively. Panel (e) exhibits the evolution of the band gap as a function of strain, where blue and orange color regions depict the direct and indirect band gap.}
\end{figure*}
In the low-energy Hamiltonian near the $K$ and $\bar{K}$ valleys (see \cref{eq:kpValley}), the Berry curvature is controlled by the terms $H_\text{mass}$ and $H_\text{SOI}$. Since \cref{eq:kpValley} is a two-band model Hamiltonian in the $\sigma_i$ basis, we can write it in the form $H=\vec{d}(\vec{k})\cdot\vec{\sigma}$. The term responsible for controlling the Berry curvature is the $z$-component of $\vec{d}(\vec{k})$:
\begin{equation}
    d_z= \Delta + \delta s + \lambda_is\tau + \lambda_u\tau,
\end{equation}
where $\tau$ and $s$ represents the valley and spin index each taking values $\pm 1$.
The $z$-component of the Berry curvature takes the form (see Appendix~\ref{app:BerryCurvature})
\begin{equation}
    \Omega_z^\pm(\vec{k}) = \mp \frac{1}{2}~ \frac{v_F^2 \tau d_z}{\left(v_F^2 k^2 + d_z^2\right)^{3/2}},
\end{equation}
with $\pm$ representing the band index corresponding to the conduction and valence band, respectively. A finite net Berry curvature arises only when a nonzero mass gap $d_z$ induces an out-of-plane pseudospin component. The term $\Delta$ breaks the inversion symmetry and generates an opposite Berry curvature in the two valleys, while the spin-orbit coupling term $\lambda_i s_z\tau_z$ introduces spin-valley dependence, leading to $\Omega(K) \neq - \Omega(\bar{K})$, hence a finite net Berry curvature arises, as shown in \cref{fig:inequivalent_model}(c) and \ref{fig:inequivalent_model}(d). The spin-valley coupled Berry curvature in this system originates from the spin- and valley-dependent $d_z$ term in the effective Hamiltonian. Due to the presence of spin-orbit coupling, the electronic states near the $K$ and $\bar{K}$ valleys become intrinsically spin-polarized, leading to a strong coupling between spin and valley degrees of freedom. As a result, the Berry curvature becomes explicitly dependent on both spin and valley indices \cite{FengPRB12}.
\subsection{Effects of Strain}
We have considered an in-plane biaxial strain to investigate the effects on the electronic properties of the MnPS$_3|$WS$_2$ heterostructure. The variation of interlayer distance as a function of strain is shown in \cref{fig:strain}(a). Since strain affects the band dispersion in a qualitatively similar manner for all three stacking configurations, we present and discuss only the DFT results for the stacking configuration II. For strain varying from $-6\%$ to $7\%$, \cref{fig:strain}(a) shows a continuous decrease of the interlayer gap within the range of $\sim \pm 0.03$~\AA\ from the equilibrium structure. The evolution of the band gap and the relative movement of the valence bands near the $\Gamma$-point in response to strain are demonstrated in \cref{fig:strain}(b), \ref{fig:strain}(c), and \ref{fig:strain}(d). The band dispersion from our DFT results suggests that compressive strain decreases the band gap while preserving a direct band gap at $K$ and $\bar{K}$ points, while excessive tensile strain shifts the direct band gap at the $\Gamma$-point. Upon the application of compressive strain, the locations of the band edge positions relative to the Fermi level remain unchanged. Hence, the band alignment changes from type II to type I. We observe a direct band gap for almost the entire spectrum of strain, except for a small region, as schematically illustrated in \cref{fig:strain}(e). We note from \cref{fig:strain}(f), \ref{fig:strain}(g), and \ref{fig:strain}(h) that at the $\Gamma$ point, VBM is contributed by W $d_{3z^2-r^2}$ and S $p_{z}$ states, responsible for Rashba-like splitting; the in-plane strain up to $\varepsilon=-2\%$ sustains this overlapping at the $\Gamma$ point, but with increasing compressive strain, contribution from S $p$ only remains at the $\Gamma$ point. Hence, due to the lack of overlap between responsible states, the Rashba-like splitting at the VBM vanishes. Whereas with the in-plane strain of $\varepsilon=-4\%$, the WS$_2$ bands move downward and become the CBM, and MnPS$_3$ bands become higher than the WS$_2$ layer, indicating the transformation from type II to type I band alignment and direct band gap at $K$/$\bar{K}$.
\begin{figure}
\includegraphics[scale=0.20]{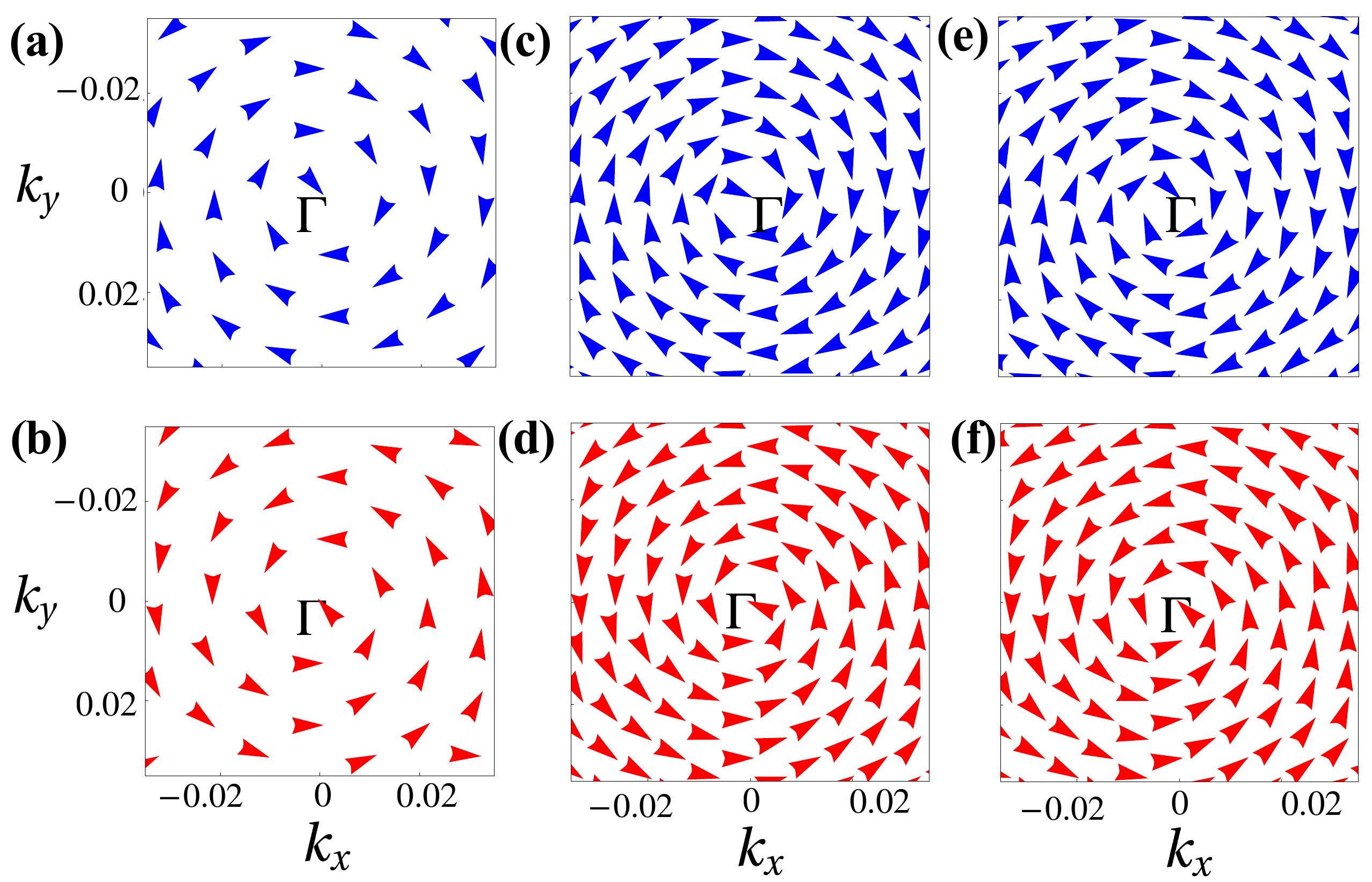}
\caption{\label{fig:Rashba_strain} The spin-textures for the highest two valence bands at $\Gamma$-point after being obtained from the DFT results applying strain presented here. The spin textures of Rashba-like splitting in the VBM from our DFT calculations for the constant energy are displayed in panels (a),(b) $2\%$, (c),(d) $4\%$, and (e),(f) $7\%$, for the set of pairs of bands marked in \cref{fig:strain}(a).}
\end{figure}
In contrast, the change in the band gap for the heterostructure is non-monotonic under in-plane tensile strain. The band gap initially increases up to $\varepsilon = 4\%$, and then, beyond $4\%$, it decreases drastically. Furthermore, the transition of the band gap from direct at the $K$/$\bar{K}$ point to an indirect one ($\varepsilon=5\%$) and subsequently back to a direct band gap again ($\varepsilon = 6\%$) \cite{YaoPRB17}. An increase in in-plane tensile strain ($\varepsilon = 4\% \text{ to } 7\%$) reduces the overlap of $d_{3z^2-r^2}$ and $p_z$ states, leading to antibonding states rising in the VBM and weak hybridization lowering the bonding states in the CBM at the $\Gamma$ point. This results in the formation of a direct band gap, with the decrease in band gap at the $\Gamma$ point. It should be noted that, upon applying the tensile strain, the overlap of the W $d_{x^2-y^2}$ + $d_{xy}$ and S $p_{x}$ + $p_{y}$ antibonding states increases in the K-point of VBM, which accounts for the observable spin splitting in the $K$/$\bar{K}$ point. Conversely, the introduction of compressive strain suppresses the coupling between analogous antibonding states, thereby reducing spin splitting in the bands. The spin textures corresponding to the pair of spin-split valence bands at the $\Gamma$-point, subject to $-4$\%, 4\%, and 7\% strain, are presented in \cref{fig:Rashba_strain}(a,b), \ref{fig:Rashba_strain}(c,d), and \ref{fig:Rashba_strain}(e,f), respectively. The observed helical spin textures confirm a Rashba-type SOI, whose strength is enhanced with an increase in tensile strain. These findings demonstrate that in-plane strain can effectively modulate the electronic structure of the MnPS$_3 | $ WS$ _ 2$ heterostructure, including the band gap, band alignment, and spin splitting, and will be utilized to design various nanodevices.
\subsection{Effect of External Electric Field}
\begin{figure}
\includegraphics[scale=0.175]{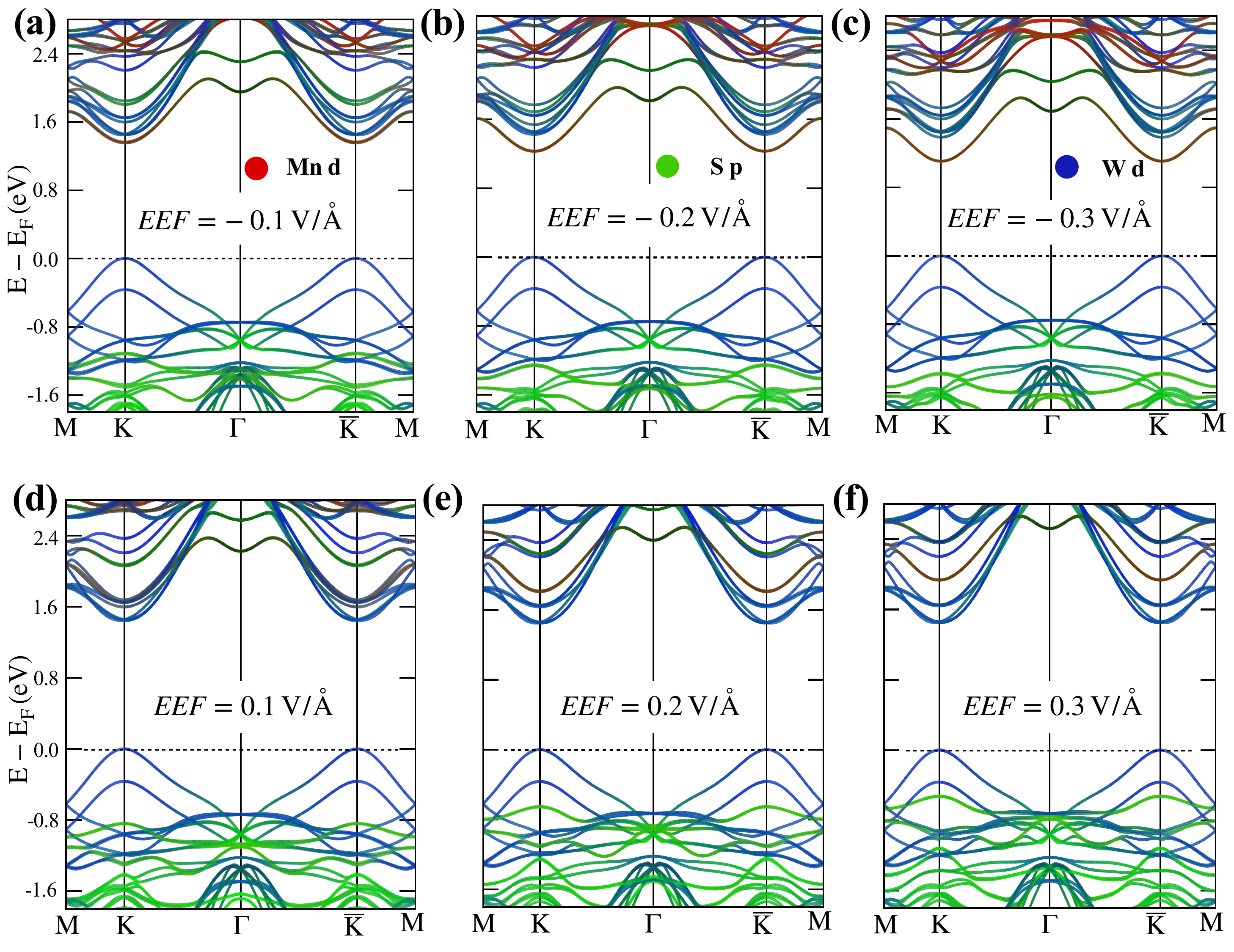}
\caption{\label{fig:ExternalEF}The electric field dependent band splitting of the MnPS$_3|$WS$_2$ heterostructure including SOI, is shown along the high-symmetry path $M \rightarrow K \rightarrow \Gamma \rightarrow \bar{K} \rightarrow M$. Panels (a)-(c) display the layer-projected band structure under a negative EEF, while panels (d)-(f) present the corresponding results for a positive EEF applied along the stacking direction of the heterostructure. Here, the positive and negative signs denote the polarity of the applied electric field.}
\end{figure}

\begin{figure*}
 \includegraphics[scale=0.28]{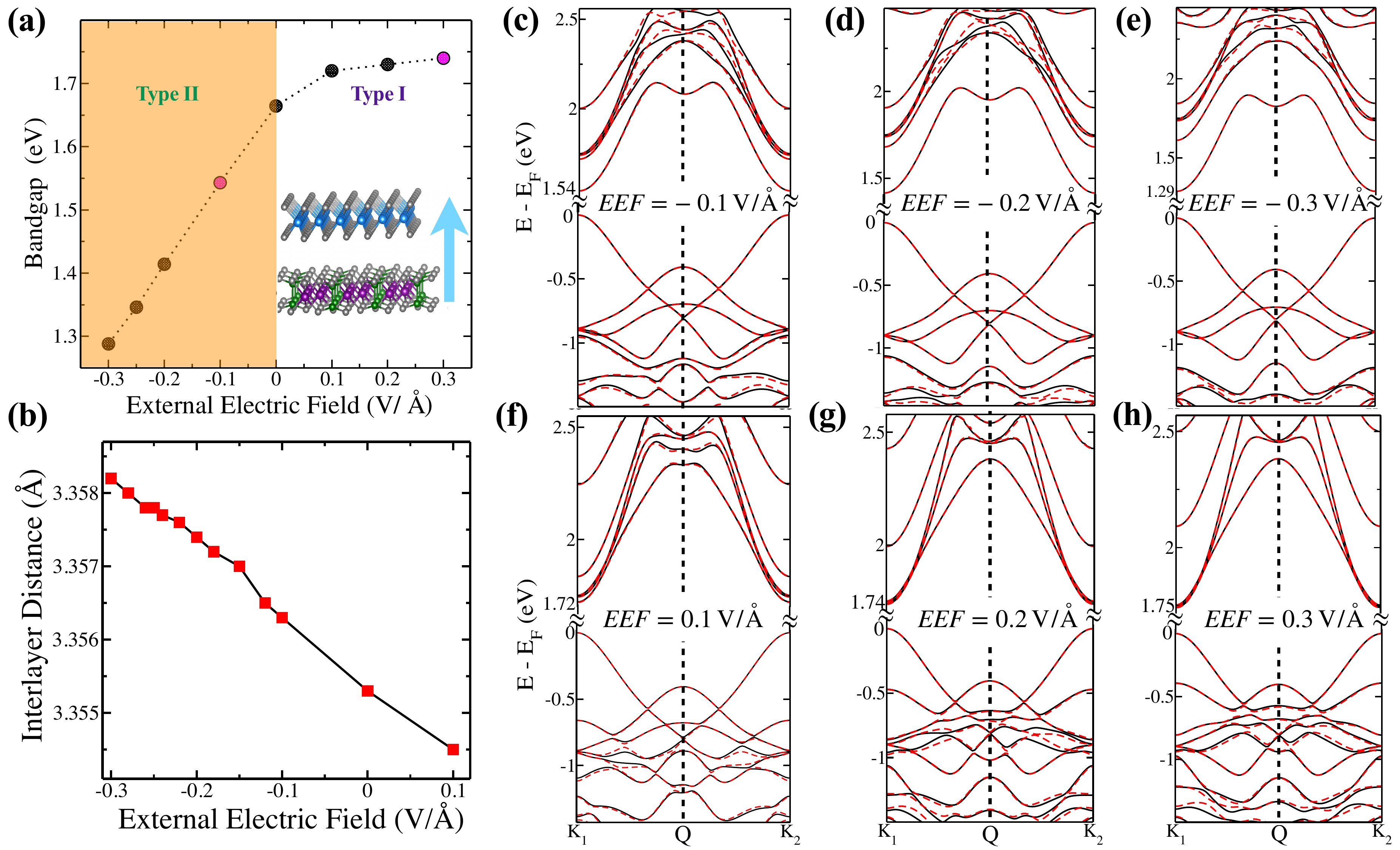}
 \caption{\label{fig:alt_EEF}The variation of band alignment, interlayer distance, and electric field induced band splitting in the MnPS$_3|$WS$_2$ heterostructure, without SOI, is shown as a function of the EEF along the high-symmetry direction $K_1 \to K_2$. Panel (a) presents the evolution of the band gap under both positive and negative EEF, where the blue shaded region denotes type-II band alignment, and the white region corresponds to type-I alignment. Panel (b) illustrates the corresponding modulation of the interlayer distance with EEF. Panels (c)-(e) display the enhancement in splitting of the conduction bands under negative EEF, whereas panels (f)-(h) show the suppression of the lifted degeneracy in the conduction bands under positive EEF applied along the stacking direction of the heterostructure. Here, the positive and negative signs indicate the polarity of the applied electric field.}
\end{figure*}

The vertically stacked MnPS$_3|$WS$_2$ heterostructure experiences a built-in electric field across the interface due to the potential difference between the layers (see \cref{fig:Stacking}(g)), leading to novel properties absent in conventional layers. The strength of the intrinsic electric field can be augmented or diminished by application of an external electric field (EEF). An EEF from MnPS$_3$ to WS$_2$, perpendicular to the interface, is considered as a positive field. Such an EEF alters the orbital overlap at the interface, thereby modifying the band alignment of these hybridized states. While an increasing positive EEF retains the type-I nature of the band alignment, an increasingly negative EEF changes the alignment to type-II by enhancing the hybridization between Mn $d$ and S $p$ states \cite{PatelPRB22, XiaPRAp18}, as shown in \cref{fig:ExternalEF}(a)–(c). Conversely, a positive EEF suppresses this Mn $d$ and S $p$ hybridization by pushing these states to higher energy, preserving the type-I band alignment, as shown in \cref{fig:ExternalEF}(d)–(f).
\Cref{fig:alt_EEF}(a) and (b) reveal the variation in band gap and monotonically decrease with the interlayer separation as the EEF turns from negative to positive values. An increasingly negative EEF tends to decrease the band gap and shift the MnPS$_3$ bands towards the lowest energy level, as demonstrated in \cref{fig:alt_EEF}(c), \ref{fig:alt_EEF}(d), and \ref{fig:alt_EEF}(e)). Conversely, an increasingly positive EEF tends to slightly increase the band gap and shift the WS$_2$ bands downward, as evident from \cref{fig:alt_EEF}(f), \ref{fig:alt_EEF}(g), and \ref{fig:alt_EEF}(h). However, the characteristic of the band gap remains direct in both cases of EEF at the high symmetry $\vec{k}$-point. Additionally, from the band dispersions without SOI obtained from our DFT calculations, we observe that along the high-symmetry direction $K_1 \rightarrow K_2$, the conduction bands associated with the sublattices exhibit a slight enhancement in splitting up to $E \approx -0.2$~eV/\AA\, when an EEF is applied opposite to the intrinsic electric field. Beyond this energy range, the splitting is gradually reduced. In contrast, when the EEF is applied along the direction of the intrinsic electric field, the splitting in the conduction bands is suppressed, eventually vanishing along the $K_1 \rightarrow K_2$ path in the Brillouin zone.

\begin{figure*}
\includegraphics[scale=0.26]{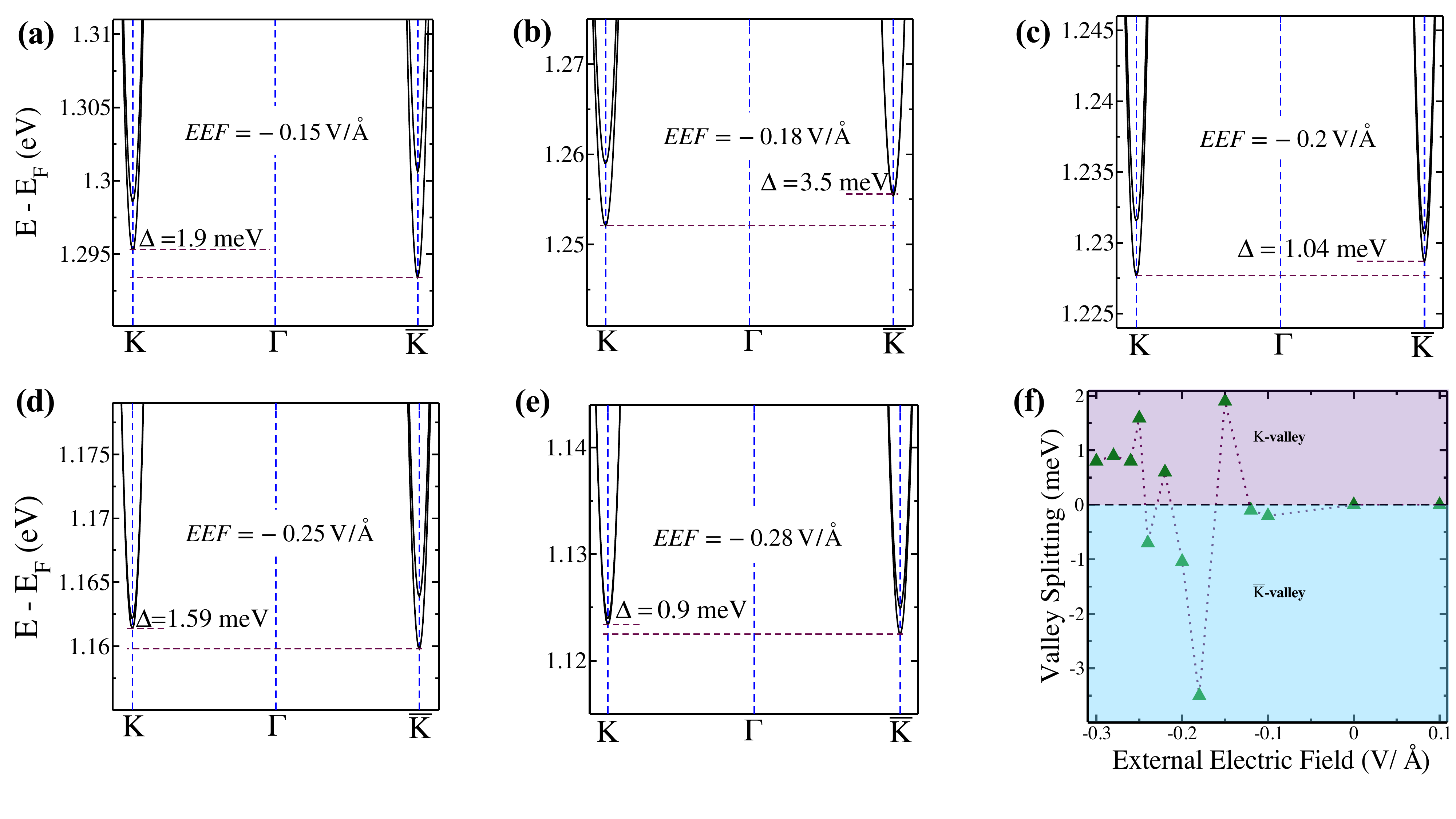}
\caption{\label{fig:valley}The variation of band dispersion near the valley points with the applied electric field in the momentum space and switching of valley polarization at the various EEF is presented here. Panels (a), (d), and (e) show the variation of valley splitting corresponding to the applied electric field at the $K$-valley, while panels (b)and (c) show valley splitting corresponding to the $\bar{K}$-valley. The evolution of valley-split gap as a function of EEF is demonstrated in panel (f), where purple and blue colors depict the splitting at the $K$-valley and $\bar{K}$-valley regions, respectively.}
\end{figure*}

These findings reveal that the electric field is an effective knob for tuning the altermagnetic splitting in the conduction bands and the band alignment. The effect of the EEF on the Rashba-like splitting in the valence bands at the $\Gamma$-point is minimal under negative field conditions, while increasing the positive EEF further suppresses the spin splitting. Reversing the electric field from positive to negative increases the interlayer distance between the MnPS$_3$ and WS$_2$ layers, thereby weakening interfacial orbital hybridization and reducing interlayer coupling, and changes the bond lengths relative to the zero-field configuration. These observations collectively indicate a field-induced redistribution of charge at the interface upon reversal of the electric field polarity. We next investigated the effect of EEF on the $K$ and $\bar{K}$ valleys in the type-II stacking configuration, focusing on the field dependence of the valley splittings. With the detailed inspection of the band dispersion with SOI, we find that the valley splitting appears at the valley points $K$ and $\bar{K}$ with the negative EEF, and the evolution of the valley splitting occurs with a critical negative EEF. However, upon reaching a critical EEF, the valley splitting diminished at the $K$ or $\bar{K}$ valley, as shown in \cref{fig:valley}(a)-(e). The evolution of $\Delta$ is clearly illustrated in \cref{fig:valley}(f) and interestingly, is non-monotonic but exhibits several sharp discontinuities. The valley splitting at $K$ and $\bar{K}$ for different electric fields reveals that the maximum valley splitting of 3.5~meV is observed at $EEF=-0.18$~eV/\AA. The evolution of valley splitting with EEF is governed by a complex interplay of symmetry breaking, band alignment, and orbital character near the valley edges. While the electric field enhances inversion asymmetry and initially increases the splitting, further increase in field strength leads to significant modifications in the electronic structure, including band reordering and changes in orbital contributions. These effects result in a non-monotonic variation of the valley splitting, with sharp features arising from transitions in the dominant orbital character. The role of interlayer hybridization, which is also modified in the applied field, provides an additional contribution to this behavior.
\section{\label{sec:summary}Conclusions}
In this study, we employ density functional theory to investigate the structural, electronic, and magnetic properties of monolayer MnPS$_3$, WS$_2$, and their vdW heterostructures. By systematically applying both strain and out-of-plane electric fields, we elucidate how external perturbations govern the interplay between interlayer interactions, spin-orbit coupling, and magnetic ordering. Owing to the weak vdW bonding at the interface, the intrinsic electronic characteristics of MnPS$_3$ and WS$_2$ remain largely preserved within the heterostructure. The layer stacking introduces interfacial distortion in the local sublattice environments, breaking $\mathcal{PT}$ symmetry, while preserving the rotational symmetry that connects the two magnetic sublattices, unveiling altermagnetic signatures that are absent in the MnPS$_3$ monolayer. These altermagnetic signatures persist over a wide range of electric-field strengths, demonstrating the robustness of the magnetic configuration against external fields. The band-edge states show type-I band alignment, with the CBM and the VBM residing primarily in WS$_2$. In addition, we observe a pronounced Rashba spin splitting in the valence bands near the Fermi level at the $\Gamma$ point, originating from the combined influence of interfacial spin-orbit coupling and inversion-symmetry breaking. Our systematic and critical analysis of the spin-orbit interaction through the band dispersion reveals valley-dependent energy splitting at the  $K$ and $\bar{K}$ valley points. The observed splitting arises from the stacking-induced anisotropy of the local crystal environment, which breaks the local inversion symmetry at the valley sites and activates the intrinsic spin-orbit coupling. We have also shown that the electronic response to an external electric field is highly sensitive to the interfacial stacking environment between the MnPS$_3$ and WS$_2$ layers. The applied electric field induces subtle distortions in the interlayer geometry, providing insight into field-driven modification of the structural landscape. These distortions lead to measurable valley splitting in the antiferromagnetic state, revealing a strong coupling between magnetic order, lattice asymmetry, and valley degrees of freedom. Furthermore, we discover that tensile strain serves as an efficient method for manipulating the band alignment characteristics and the manifestation of spin-orbit interaction. Increasing tensile strain enhances the Rashba splitting strength and enables controllable transitions between band gap types, thereby providing an alternative route for tuning the electronic properties beyond the capabilities of electric fields. This strain-driven modulation demonstrates the adaptability of the MnPS$_3|$WS$_2$ interface for spin-orbit engineering and bandstructure control.
Finally, the band gap exhibits strong sensitivity to the external electric field, allowing for reversible switching between type-I and type-II band alignments. These results emphasize that the MnPS$_3|$WS$_2$ heterostructure is a promising platform for spintronic, valleytronic, and optoelectronic functionalities, where the electronic and magnetic responses can be effectively tuned via electric fields and mechanical strain.\\


\begin{acknowledgments}
P.D.\ would like to acknowledge UGC, India, for the research fellowship [grant no.1498/(CSIR NET JUNE 2019)] and the use of the high-performance computing facility (HPC) at IISER Bhopal.
\end{acknowledgments}

\appendix
\section{Derivation of Berry Curvature for a Two-Band Hamiltonian}
Here, we derive an expression for the $z$-component of the Berry curvature $\Omega_z^\pm(\vec{k})$ of a generic two-band Hamiltonian $H = \vec{d}(\vec{k}) \cdot \vec{\sigma}$ having eigenstates as $|u_{\pm}(\vec{k})\rangle$ and energy eigenvalue $E_{\pm} = \pm |\vec{d}|$:
\begin{equation}
\Omega_z^\pm(\vec{k}) = i \left(\langle \partial_x u_{\pm} |\partial_y u_{\pm} \rangle-\langle \partial_y u_{\pm} |\partial_x u_{\pm} \rangle\right),
\label{eq:A1}
\end{equation}
where $\partial_i \equiv \partial/\partial k_i$.
Using the completeness relation,
\begin{equation}
I = |u_{\pm}\rangle\langle u_{\pm}| + |u_{\mp}\rangle\langle u_{\mp}|,
\label{eq:A2}
\end{equation}
we obtain
\begin{align}
\Omega_z^\pm(\vec{k}) =& i \langle \partial_x u_{\pm} | u_{\pm} \rangle \langle u_{\pm} | \partial_y u_{\pm} \rangle \nonumber \\
& + i \langle \partial_x u_{\pm} | u_{\mp} \rangle\langle u_{\mp}|\partial_y u_{\pm} \rangle \nonumber \\
& -i \langle \partial_y u_{\pm} | u_{\pm} \rangle \langle u_{\pm} | \partial_x u_{\pm} \rangle \nonumber \\
& -i\langle \partial_y u_{\pm} | u_{\mp} \rangle \langle u_{\mp} | \partial_x u_{\pm} \rangle .
\label{eq:A3}
\end{align}
The first and third terms correspond to Berry connection contributions and can be eliminated by an appropriate choice of gauge, leading to
\begin{align}
 \Omega_z^\pm(\vec{k})=& i[\langle \partial_x u_{\pm} | u_{\mp} \rangle\langle u_{\mp} | \partial_y u_{\pm} \rangle \nonumber \\
&- \langle \partial_y u_{\pm} | u_{\mp} \rangle\langle u_{\mp} | \partial_x u_{\pm} \rangle ].
\label{eq:A4}
\end{align}
Using the well-known Hellmann-Feynman identity \cite{Bransden}
\begin{equation}
\langle u_{\mp} | \partial_i u_{\pm} \rangle = \frac{\langle u_{\mp} | \partial_i H | u_{\pm} \rangle}{E_\pm - E_{\mp}},
\label{eq:A5}
\end{equation}
we obtain the following expression for Berry curvature
\begin{align}
    \Omega_z^\pm(\vec{k})= \frac{i}{4|\vec d|^2}[&\langle u_{\pm} | \partial_x H | u_{\mp} \rangle \langle u_{\mp} | \partial_y H | u_{\pm} \rangle \nonumber \\
    &-\langle u_{\pm} | \partial_y H | u_{\mp} \rangle \langle u_{\mp} | \partial_x H | u_{\pm} \rangle].
\label{eq:A6}
\end{align}

\section{\label{app:BerryCurvature}Evaluation of the matrix element}
In this section, we evaluate the matrix elements of \cref{eq:A6} one by one.
Define
\begin{align}
    B=&\langle u_{\pm}|\partial_x H|u_{\mp}\rangle \langle u_{\mp}|\partial_y H|u_{\pm}\rangle \nonumber \\  & -\langle u_{\pm}|\partial_y H|u_{\mp}\rangle \langle u_{\mp}|\partial_x H|u_{\pm}\rangle.
\label{eq:A7}
\end{align}
Using the completeness identity \cref{eq:A2}, we obtain
\begin{align}
    B =& \langle u_{\pm}|\partial_x H (I-|u_{\pm}\rangle\langle u_{\pm}|) \partial_y H|u_{\pm}\rangle \nonumber \\ & -\langle u_{\pm}|\partial_y H (I-|u_{\pm}\rangle\langle u_{\pm}|) \partial_x H|u_{\pm}\rangle .
\label{eq:A9}
    \end{align}
For a two-band Hamiltonian, the projection operator can be written as
\begin{equation}
|u_{\pm}\rangle\langle u_{\pm}|=\frac{1}{2} \left(I \pm \frac{\vec d\cdot\vec{\sigma}}{|\vec d|}\right).
\label{eq:A10}
\end{equation}
Substituting this into the expression \cref{eq:A9} yeilds
\begin{align}
B &= \frac{1}{2} \langle u_{\pm}|\partial_x H \partial_y H|u_{\pm}\rangle
\mp \frac{1}{2|\vec d|}\langle u_{\pm}|\partial_x H(\vec{\sigma}\cdot\vec d)\partial_y H|u_{\pm}\rangle \nonumber \\
& - \frac{1}{2} \langle u_{\pm}|\partial_y H \partial_x H|u_{\pm}\rangle
\pm \frac{1}{2|\vec d|}\langle u_{\pm}|\partial_y H(\vec{\sigma}\cdot\vec d)\partial_x H|u_{\pm}\rangle.
\label{eq:A11}
\end{align}
Since, the derivative with respect to $k_i$ only acts on $\vec{d}$,
\begin{equation}
\partial_i H=(\partial_i\vec d)\cdot\vec{\sigma},
\label{eq:A12}
\end{equation}
the first and third term of \cref{eq:A11} becomes
\begin{equation}
B_{1,3}=\frac{1}{2} \langle u_{\pm}|(\partial_x\vec d \cdot \vec{\sigma})
(\partial_y\vec d \cdot \vec{\sigma}) - (\partial_y\vec d \cdot \vec{\sigma})
(\partial_x\vec d \cdot \vec{\sigma})|u_{\pm}\rangle.
\label{eq:A13}
\end{equation}
Using the Pauli matrix identity
\begin{equation}
(\vec{\sigma}\cdot\vec a)(\vec{\sigma}\cdot\vec b)=\vec a\cdot\vec b+i\vec{\sigma}\cdot(\vec a\times\vec b),
\label{eq:A14}
\end{equation}
we obtain
\begin{equation}
    (\partial_x\vec d \cdot \vec{\sigma})(\partial_y\vec d \cdot \vec{\sigma}) - (\partial_y\vec d \cdot \vec{\sigma})(\partial_x\vec d \cdot \vec{\sigma}) = 2i \vec{\sigma}\cdot (\partial_x\vec d\times\partial_y\vec d).
\label{eq:A15}
\end{equation}
Using the expectation value,
\begin{equation}
\langle u_{\pm}|\vec{\sigma}|u_{\pm}\rangle = \pm\hat{d} = \pm \frac{\vec{d}}{| \vec{d} |},
\label{eq:A16}
\end{equation}
\cref{eq:A13} becomes
\begin{equation}
B_{1,3} = \pm i~\hat{d}\cdot(\partial_x\vec d\times\partial_y\vec d).
\label{eq:A17}
\end{equation}
The second and fourth term in $B$ (\cref{eq:A11}) is
\begin{equation}
B_{2,4} = \mp \frac{1}{2|\vec{d}|} \langle u_{\pm} | \{\partial_x H (\vec{\sigma} \cdot \vec{d}) \partial_y H - \partial_y H (\vec{\sigma} \cdot \vec{d}) \partial_x H\} | u_{\pm} \rangle.
\label{eq:A18}
\end{equation}
We first expand $\partial_i H = (\partial_i \vec{d}) \cdot \vec{\sigma}$:
\begin{align}
    B_{2,4} =& \mp \frac{1}{2|\vec{d}|} \langle u_{\pm} | \{(\partial_x \vec{d} \cdot \vec{\sigma}) (\vec{\sigma} \cdot \vec{d}) (\partial_y \vec{d} \cdot \vec{\sigma}) \nonumber \\
    &-(\partial_y \vec{d} \cdot \vec{\sigma}) (\vec{\sigma} \cdot \vec{d}) (\partial_x \vec{d} \cdot \vec{\sigma})\} | u_{\pm} \rangle.
 \label{eq:A19}
\end{align}
Using the Pauli matrix identity, we first evaluate the product
$(\vec{\sigma}\cdot \vec{d})(\partial_y \vec{d}\cdot \vec{\sigma})$ as
\begin{equation}
(\vec{\sigma}\cdot \vec{d})(\partial_y \vec{d}\cdot \vec{\sigma}) = \vec{d}\cdot \partial_y \vec{d} + i \vec{\sigma}\cdot (\vec{d} \times \partial_y \vec{d}).
\label{eq:A20}
\end{equation}
Thus, $B_{2,4}$ becomes
\begin{align}
B_{2,4} =& \mp \frac{1}{2|\vec{d}|} \langle u_{\pm} | \{(\partial_x \vec{d}\cdot \vec{\sigma}) \left[ \vec{d}\cdot \partial_y \vec{d} + i \vec{\sigma}\cdot (\vec{d} \times \partial_y \vec{d}) \right] \nonumber \\
& - (\partial_y \vec{d}\cdot \vec{\sigma}) \left[ \vec{d}\cdot \partial_x \vec{d} + i \vec{\sigma}\cdot (\vec{d} \times \partial_x \vec{d}) \right] \}| u_{\pm} \rangle \nonumber \\
=& \mp \frac{1}{2|\vec{d}|} \Big[ \langle u_{\pm} | \partial_x \vec{d}\cdot \vec{\sigma} | u_{\pm} \rangle (\vec{d}\cdot \partial_y \vec{d}) \nonumber \\
& + i \langle u_{\pm} | (\partial_x \vec{d}\cdot \vec{\sigma}) (\vec{\sigma}\cdot (\vec{d} \times \partial_y \vec{d})) | u_{\pm} \rangle \nonumber   \\
& -\langle u_{\pm} | \partial_y \vec{d}\cdot \vec{\sigma} | u_{\pm} \rangle (\vec{d}\cdot \partial_x \vec{d}) \nonumber \\
& - i \langle u_{\pm} | (\partial_y \vec{d}\cdot \vec{\sigma}) (\vec{\sigma}\cdot (\vec{d} \times \partial_x \vec{d})) | u_{\pm} \rangle \Big].
\label{eq:A21}
\end{align}
Now, using the fact that
\begin{equation}
\langle u_{\pm} | \vec{\sigma} | u_{\pm} \rangle = \pm \hat{d} \implies \langle u_{\pm} | \partial_x \vec{d}\cdot \vec{\sigma} | u_{\pm} \rangle = \pm \hat{d}\cdot \partial_x \vec{d},
\label{eq:A22}
\end{equation}
the first term of \cref{eq:A21} cancels with the third term under $(x \leftrightarrow y)$:
\begin{equation}
\langle u_{\pm} | \partial_x \vec{d}\cdot \vec{\sigma} | u_{\pm} \rangle (\vec{d}\cdot \partial_y \vec{d}) - \langle u_{\pm} | \partial_y \vec{d}\cdot \vec{\sigma} | u_{\pm} \rangle (\vec{d}\cdot \partial_x \vec{d}) = 0.
\label{eq:A23}
\end{equation}
For the second term of \cref{eq:A21}, applying the Pauli identity again, we get
\begin{align}
(\partial_x \vec{d}\cdot \vec{\sigma}) (\vec{\sigma}\cdot (\vec{d}\times \partial_y \vec{d})) 
=& \partial_x \vec{d} \cdot (\vec{d} \times \partial_y \vec{d}) \nonumber \\
&+ i \vec{\sigma} \cdot (\partial_x \vec{d} \times (\vec{d} \times \partial_y \vec{d})).
\label{eq:A24}
\end{align}
Similarly, we apply the Pauli identity for the fourth term of \cref{eq:A21}, and combining the second and fourth terms of \cref{eq:A21}, we get,
\begin{align}
   =& [ i \langle u_{\pm} | \partial_x \vec{d} \cdot (\vec{d} \times \partial_y \vec{d}) + i \vec{\sigma} \cdot (\partial_x \vec{d} \times (\vec{d} \times \partial_y \vec{d})) | u_{\pm} \rangle \nonumber \\
    & - i \langle u_{\pm} | \partial_y \vec{d} \cdot (\vec{d} \times \partial_x \vec{d}) + i \vec{\sigma} \cdot (\partial_y \vec{d} \times (\vec{d} \times \partial_x \vec{d})) | u_{\pm} \rangle ] \nonumber \\
    =& \big[ 2i \langle u_{\pm} | \vec{d} \cdot (\partial_y \vec{d} \times \partial_x \vec{d}) | u_{\pm} \rangle \nonumber \\
    & \pm \hat{d} \underbrace{[\partial_y \vec{d} \times (\vec{d} \times \partial_x \vec{d}) - \partial_x \vec{d} \times (\vec{d} \times \partial_y \vec{d})]}_{0} \big ]
\end{align}
This leaves with the final form of $B_{2,4}$ as
\begin{equation}
B_{2,4} = \pm \frac{1}{2|\vec{d}|} 2 i \vec{d} \cdot (\partial_x \vec{d} \times \partial_y \vec{d}) 
=  \pm i~\hat{d} \cdot (\partial_x \vec{d} \times \partial_y \vec{d}).
\label{eq:A25}
\end{equation}
Adding the contributions from $B_{1,3}$ and $B_{2,4}$, we finally obtain
\begin{align}
    \Omega_z^\pm(\vec{k}) 
&= \frac{i}{4|\vec{d}|^2} (B_{1,3}+B_{2,4}) \nonumber \\
&=\pm \frac{i}{4|\vec{d}|^2} 2 i \hat{d} \cdot (\partial_x \vec{d} \times \partial_y \vec{d}) \nonumber \\
&= \mp \frac{1}{2} \frac{\vec{d} \cdot (\partial_x \vec{d} \times \partial_y \vec{d})}{|\vec{d}|^3}.
\label{eq:A26}
\end{align}
This is the Berry curvature for a general two-band Hamiltonian.

%
\end{document}